\documentclass[12pt]{article}
\pdfoutput=1
\usepackage{slashed}
\usepackage{amsmath,amssymb}
\usepackage{bbm}
\usepackage{cite}
\usepackage[bookmarks=false]{hyperref}

\usepackage{graphicx}
\usepackage{amsbsy}
 
\newcommand{\beq}{\begin{equation}}
\newcommand{\eeq}{\end{equation}}
\newcommand{\bea}{\begin{eqnarray}}
\newcommand{\bi}{\begin{itemize}}
\newcommand{\eea}{\end{eqnarray}}
\newcommand{\ei}{\end{itemize}}
\newcommand{\3}{\mbox{${\bf \underline{3}}$}}
\newcommand{\s}{\mbox{${\bf \underline{1}}$}}
\newcommand{\spr}{\mbox{${\bf \underline{1}'}$}}
\newcommand{\sppr}{\mbox{${\bf {\underline{1}''}}$}}
\begin{document}
\setlength{\unitlength}{1mm}
\begin{titlepage}
\hypersetup{pageanchor=false}
\vspace*{0.8cm}
\begin{center}
 {\huge \bf   $\theta_{13}$ and charged lepton flavor violation in ``warped" A$_4$ models }
\end{center}
\vskip0.2cm

\begin{center}
 {\bf Avihay Kadosh$^a$ }
\end{center}
\vskip 4pt

\begin{center}
$^a$ {\it Centre for Theoretical Physics, University of Groningen,
9747 AG, Netherlands}

\vspace*{0.1cm}

{\tt a.kadosh@rug.nl }
\end{center}
\vglue 0.3truecm

\begin{abstract}
\vskip 3pt \noindent

We recently proposed a spontaneous A$_{4}$ flavor symmetry
breaking scheme implemented in a warped extra dimensional setup to
explain the observed pattern of quark and lepton masses and
mixings. The main features of this choice are the explanation of
fermion mass hierarchies by wave function overlaps, the emergence
of tribimaximal (TBM) neutrino mixing and zero quark mixing at the
leading order and the absence of tree-level gauge mediated flavor
violation. Quark mixing and deviations from \,TBM neutrino mixing
are induced by the presence of bulk A$_4$ flavons, which allow for
``cross-brane'' interactions and a ``cross-talk'' between the
quark and neutrino sectors.

% We subsequently studied the
%constraints coming from electroweak precision measurements 
%and flavor and CP violating observables in the quark sector,
%demonstrating the role of A$_4$ in ameliorating the "little CP
%problem" characteristic of flavor anarchic setups.

In this work, we study the constraints associated with the recent
measurements of $\theta_{13}\approx 9^\circ$ by RENO and Daya Bay,
forcing every model that predicts TBM neutrino mixing to account
for the significant deviation of $\theta_{13}$ from 0, while
keeping the values of $\theta_{12}$ and $\theta_{23}$ close to
their central experimental values. We then proceed to study in detail the RS-A$_4$ contributions 
to $\mu\to e,3e$, generated at the tree level by virtue of anomalous $Z$
couplings. These couplings arise from gauge and fermionic KK mixing effects after electroweak symmetry breaking.
Since the experimental sensitivity for $BR(\mu\to e,3e)$ is expected to increase by  five orders of magnitude
within the next decade, it is shown that the RS-A$_4$ lepton sector can be significantly constrained. Finally, we show that
when ``cross-brane" interactions are turned off, the $Z$ couplings are protected against all anomalous contributions and a strong correlation between $\theta_{13}$ and the deviation from maximality of $\theta_{23}$ is found.

% Finally, we demonstrate that  the pasimistic case of non discovery by all future experiments, will render the ``Cross-talk"' interactions unnatural, thus forcing us to reconsider the original RS-A$_4$ brane localized setup, in which deviations of $\theta_{13}$ from zero are generated only by higher order effects on the UV brane.
\end{abstract}
\end{titlepage}
\hypersetup{pageanchor=true}

\section{Introduction}
Recently  we have proposed a model \cite{A4Warped}  based on a
bulk $A_4$ flavor symmetry \cite{a4,a4F} in warped geometry \cite{RS},
in an attempt to account for the  hierarchical charged fermion
masses, the hierarchical mixing pattern in the quark sector and
the large mixing angles and  mild hierarchy of masses in the
neutrino sector. In analogy with a previous RS realizations of
A$_{4}$ for the lepton sector \cite{Csaki:2008qq,JoseA4}, the
three generations of left-handed (LH) quark doublets are unified into a
triplet of $A_4$; this assignment forbids tree level  FCNCs driven
by the exchange of KK gauge bosons, as long as only brane localized effects are considered. The scalar sector of the
RS-A${}_4$ model consists of two bulk flavon fields, in addition
to a bulk Higgs field. The bulk flavons transform as triplets of
$A_{4}$, and allow for a complete
 "cross-talk" \cite{Volkas} between the $A_{4}\to Z_{2}$
spontaneous symmetry breaking (SSB) pattern associated with the
heavy neutrino sector - with scalar mediator  peaked towards the
UV brane - and the $A_{4}\to Z_{3}$ SSB pattern associated with
the quark and charged lepton sectors - with scalar mediator peaked
towards the IR brane - allowing us to obtain realistic
mixing angles in the quark sector and  deviations from TBM  mixing in the neutrino sector. A bulk
custodial symmetry, broken differently at the two branes
\cite{Agashe:2003zs}, guarantees the suppression of large
contributions to electroweak precision observables
\cite{Carena:2007}, such as the Peskin-Takeuchi $S$, $T$
parameters. However, the mixing  between zero modes of the 5D
theory and their Kaluza-Klein (KK) excitations -- after 4D
reduction -- may still cause significant new physics (NP)
contributions to SM suppressed flavor changing neutral current
(FCNC) processes. 

In \cite{A4CPV} we have performed a thorough study of the RS-A$4$ contributions
to one loop $\Delta F=1$ dipole transitions (nEDM, $\epsilon'/\epsilon_K$, $b\to s(d)\gamma$) and tree level Higgs mediated
FCNC $\Delta F=2$ transitions $(K^{0}-\bar{K^{0}}$ and $D^{0}-\bar{D^{0}} {\rm mixing})$.  
%In particular, we showed that the 
%constraints on the KK mass scale coming from the neutron EDM and $\epsilon_K$, are significantly relaxed compared to flavor anarchic schemes. The most relevant constraint was shown to come from $b\to s\gamma$ but is less significant than the 
The main difference between the RS-A$_4$ setup and an anarchic RS
flavor scheme \cite{Agashe:2004cp} lies in the degeneracy of
fermionic LH bulk mass parameters, which implies the universality
of LH zero mode profiles and hence forbids gauge mediated FCNC
processes at tree level, including the KK gluon exchange
contribution to $\epsilon_K$. The latter provides the most
stringent constraint on flavor anarchic models, together with the
neutron EDM (nEDM) \cite{Agashe:2004cp,IsidoriPLB}. However, the choice
of the common LH bulk mass parameter, $c_q^L$ is strongly
constrained by the matching of the top quark mass
($m_t(1.8$\,TeV)$\approx140$\,TeV) and the perturbativity bound of
the 5D top Yukawa coupling, $y_t$. Most importantly, when
considering the tree level corrections to the $Zb\bar{b}$ coupling
against the stringent electroweak precision measurements (EWPM) at the Z pole, we
showed \cite{A4Warped} that for an IR scale,
$\Lambda_{IR}\simeq1.8\,$TeV and $m_h\approx 125$\,GeV, $c_q^L$ is
constrained to be larger than 0.34. Assigning $c_q^L=0.4$ and
matching with $m_t$ we obtain $|y_t|<3$, which easily satisfies the
5D Yukawa perturbativity bound. For $\Lambda_{IR}\simeq1.8\,$TeV the lightest fermionic KK mode is $\tilde{b}^{(1)}$, the 
``fake" (cusodial) $SU(2)_R$ partner of the 5D fermion, whose zero mode is identified with $t_R$, with
$M_{KK}^{\tilde{b}}\simeq 0.95$\,TeV. The second most significant
constraint on the KK mass scale comes from one-loop Higgs mediated dipole operator 
contributions to the $b\rightarrow s\gamma$ process,
$M_{KK}^{b\rightarrow s\gamma}\simeq 2.55\Lambda_{IR}^{b\to s\gamma}\gtrsim 1.3\,Y$\,TeV, where $Y$ is
the overall scale of the dimensionless 5D Yukawa coefficients. The
constraints on $M_{KK}$ coming from $\epsilon'/\epsilon_K$ (nEDM) were shown to be weaker by at least factor of 2(6) (See
\cite{A4CPV}).

In this paper we focus on the charged lepton and neutrino sectors of the RS-A$_4$ setup. We start by performing a systematic study of the RS-A$_4$ predictions for the neutrino mixing angles in light of the recent measurements of $\theta_{13}\approx9^\circ$ by the RENO \cite{RENO}, Daya Bay \cite{DayaBay} and  DOUBLECHOOZ \cite{DoubleChooz} experiments, supporting previous indications in T2K\cite{T2K} and MINOS \cite{MINOS}. It is shown that  significant deviations from TBM neutrino mixing are generated by both ``cross talk" interactions in the charged lepton sector and higher order corrections to the heavy Majorana and (light) Dirac mass matrices. The model predictions are then compared with the results of the most recent global fits \cite{Fogli,Tortola}.
Subsequently, we inspect the RS-A$_4$ contributions to the charged lepton flavor violating processes $\mu\to 3e$ and $\mu\to e$ conversion. We show that the dominant contributions come from tree-level $Z$ boson exchange diagrams, where the flavor violating $Z$ boson couplings are induced by gauge and fermion KK mixing effects. We then compare our predictions with the existing upper bounds \cite{sindrum,sindrumII} and those expected to come from a series of experiments, currently under construction in J-PARC \cite{DeeMee,COMET,PRIME}, Osaka (RNCP) \cite{MuSIC}, Fermilab \cite{Mu2e,ProX} and PSI \cite{MU3e} within the next decade. 
\par The paper is organized as follows. In Sec. \ref{Sec:Model} we summarize the most relevant features of the RS-A$_4$ setup needed for the analysis of the following sections. In Sec. \ref{Sec:NUanalyze} we study the modifications of the PMNS matrix induced by higher order effects in both the neutrino and the charged lepton sectors and perform a numerical scan to test our predictions against the experimental bounds. In Sec.~\ref{Sec:Nubrane} we specialize to the brane localized RS-A$_4$ setup. In Sec.~\ref{Sec:Zcouplings} we study the effect of gauge boson and KK fermion mixing on $Z$ couplings and obtain analytical estimations of the $Z\mu e$ coupling relevant for the $\mu\to e,3e$ processes. In Sec. \ref{Sec:Znumerical} we obtain the exact results by a numerical diagonalization of the zero+KK mass matrices and perform a scan over the parameter space of the RS-A$_4$ setup including/excluding cross-brane effects. It is shown that the  brane localized version of RS-A$_4$ stays protected from KK mixing contributions to off diagonal $Z$ couplings. We conclude in Sec.\,\ref{Sec:Conclusions}.

\section{The RS-A$_4$ model}\label{Sec:Model}
\begin{center}
\begin{figure}[t]
\includegraphics[width=16truecm]{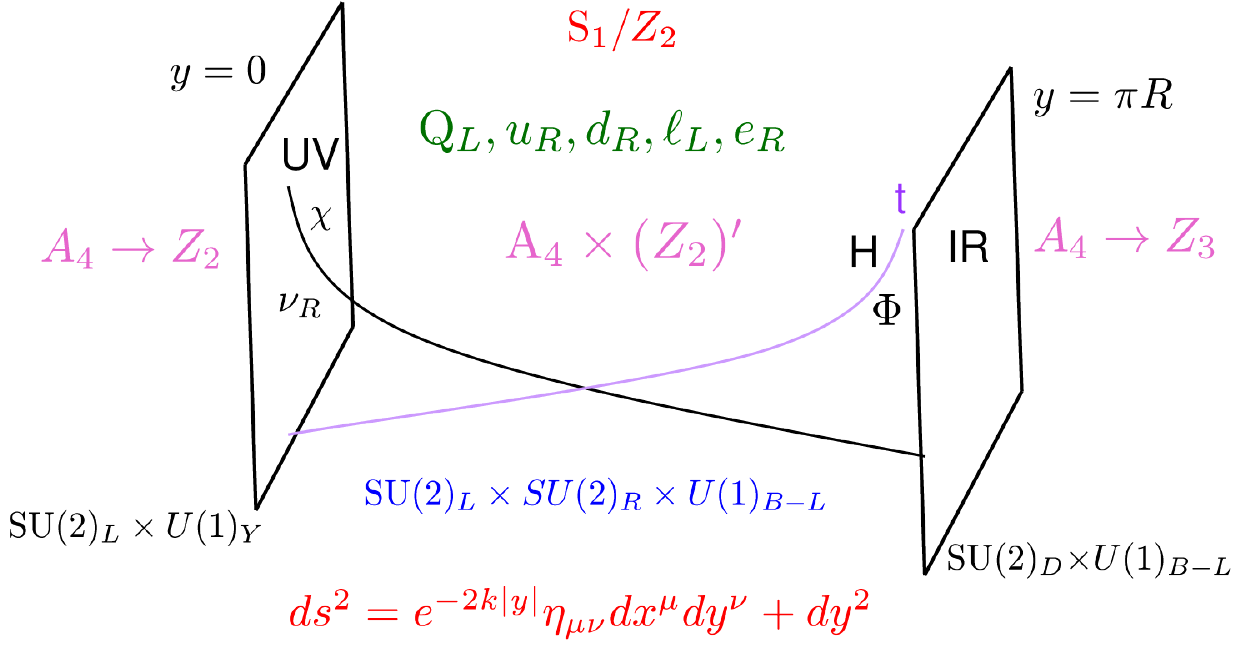}
\caption{{\small A pictorial description of the RS-A$_4$ setup.
The bulk geometry is described by the metric at the bottom and
$k\simeq M_{Pl}$ is the $AdS_5$ curvature scale. All fields
propagate in the bulk and the UV(IR) peaked nature of the heavy RH
neutrinos, the Higgs field, the $t$ quark and the A$_4$ flavons,
$\Phi$ and $\chi$, is emphasized. The SSB patterns of the bulk
symmetries on the UV and IR branes are specified on the side (for
A$_4$) and on the bottom (for $SU(2)_L\times SU(2)_R\times
U(1)_{B-L}$) of each brane. }}\label{ModelScheme}
\end{figure}
\end{center}
 The RS-A$_4$ setup  \cite{A4Warped} is illustrated  in
Fig.~\ref{ModelScheme}. The bulk geometry is that of a slice of
$AdS_5$ compactified on an orbifold $S_1/Z_2$ \cite{RS} and is
described by the metric on the bottom of Fig.~\ref{ModelScheme}.
All 5D fermionic fields propagate in the bulk and transform under
the following representations of $\left(SU(3)_c\times
SU(2)_L\times SU(2)_R\times U(1)_{B-L}\right)\times
A_4\times Z_2$\cite{A4Warped}:

\begin{equation}
\begin{array}{c}
 e_R \oplus e'_R \oplus e''_R \sim \left( 1,1,2,-1 \right)\left(\s
\oplus \spr \oplus \sppr \right)(+),\\
\\
u_R \oplus u'_R \oplus u''_R \sim \left( 3,1,2,\frac{1}{3}
\right)\left(\s \oplus
\spr \oplus \sppr \right)(+), \\
\\
d_R \oplus d'_R \oplus d''_R \sim \left( 3,1,2,\frac{1}{3}
\right)\left(\s \oplus \spr \oplus \sppr \right)(+),
\end{array}
\quad\
\begin{array}{c}
\ell_L \sim \left( 1,2,1,-1 \right) \left( \3 \right)(-), \\
\\
\nu_R \sim \left( 1,1,2,0 \right)\left( \3 \right)(-), \\
\\Q_L \sim \left( 3,2,1,\frac{1}{3} \right) \left( \3 \right)(-).
\end{array}\label{FermionContent}
\end{equation}
The SM fermions, including right handed (RH) neutrinos, are identified with the
zero modes of the 5D fermions above. The zero (and KK) mode
profiles are determined by the bulk mass of the corresponding 5D
fermion, denoted by $c_{q_L,u_i,d_i,e_i}k$ and boundary conditions (BC) \cite{A4CPV}. The
scalar sector contains the IR peaked Higgs field and the UV and IR
peaked flavons, $\chi$ and $\Phi$, respectively. They transform
as:
\begin{equation}
\Phi \sim \left( 1,1,1,0 \right) \left( \3 \right)(-),\quad \chi \sim
\left( 1,1,1,0 \right) \left( \3 \right)(+),\quad H\left(1,2,2,0
\right) \left( \s \right)(+)\, .
\end{equation}
The SM Higgs field is identified with the first KK mode of $H$.
All fermionic zero modes acquire masses through Yukawa
interactions with the Higgs field and the $A_4$ flavons after SSB.
The 5D $\left(SU(3)_c\times SU(2)_L\times SU(2)_R\times
U(1)_{B-L}\right)\times A_4\times Z_2$ invariant Yukawa Lagrangian will
consist of only UV/IR peaked interactions at the leading order (LO), while at the next to
leading order (NLO) it will also give rise to ``cross-talk" and "cross-brane"
effects  \cite{A4Warped}.
 The
LO interactions in the neutrino sector are shown in
\cite{A4Warped} using the see-saw I mechanism, to induce a
tribimaximal (TBM)\cite{TBM} pattern for neutrino mixing. Below we are going to show that 
"cross brane" and "cross talk" interactions, induce 
deviations from TBM, such that the model predictions for $\theta_{12,23,13}$  are  in good
agreement with the current experimental bounds \cite{Fogli,Tortola}.
Starting from the  quark and charged lepton sectors, the  most relevant terms
of the 5D Yukawa Lagrangian are of the following form:

\begin{equation}
\mathcal{L}_{5D}^{Y\!uk.}\supset\underbrace{\frac{y_{u_i,d_i,e_i}}{k^2}(\overline{Q}_L,\bar{\ell}_L)\Phi
H(u_R^{(\prime,\,\prime\prime)},d_R^{(\prime,\,\prime\prime)},e_R^{(\prime,\,\prime\prime)})}_{
\mathcal{L}_{LO}}
+\underbrace{\frac{(\tilde{x}_i^{u,d,\ell},\tilde{y}_i^{u,d,\ell})}{k^{7/2}}(\overline{Q}_L,\bar{\ell}_L)\Phi\chi
H(u_R^{(\prime,\,\prime\prime)},d_R^{(\prime,\,\prime\prime)},e_R^{(\prime,\,\prime\prime)})}_{\mathcal{L}_{NLO}}.
\label{LYuk5D}\end{equation}

\noindent Notice that the LO interactions are peaked towards the
IR brane while the NLO interactions mediate between the two branes
due to the presence of both $\Phi$ and $\chi$.

 \noindent The VEV and physical profiles
for the bulk scalars are obtained by solving the corresponding
equations of motion with a UV/IR localized quartic potential term
and an IR/UV localized  mass term \cite{WiseScalar}. In this way
one can obtain either UV or IR peaked and also flat profiles
depending on the bulk mass and the choice of boundary conditions.
The resulting VEV profiles of the RS-A$_4$ scalar sector are:
\begin{equation}
v_{H(\Phi)}^{5D}=H_0(\phi_0)e^{(2+\beta_{H(\phi)})k(|y|-\pi
R)}\qquad
v_\chi^{5D}=\chi_0e^{(2-\beta_\chi)k|y|}(1-e^{(2\beta_\chi)k(|y|-\pi
R)})\,,\label{VEVprofile}
\end{equation}
where $\beta_{H,\Phi,\chi}=\sqrt{4+\mu_{H,\Phi,\chi}^2}$ and
$\mu_{H,\Phi,\chi}$ is the bulk mass of the corresponding scalar
in units of $k$, the cutoff of the 5D theory. The following vacua
for the Higgs and the $A_4$ flavons $\Phi$ and $\chi$,
\begin{equation}
\langle
\Phi\rangle=(v_\phi,v_\phi,v_\phi)\qquad\langle\chi\rangle=(0,v_\chi,0)\qquad
\langle H\rangle=v_
H\left(\begin{array}{cc}1&0\\0&1\end{array}\right)\label{VEValignment},\end{equation}
provide at LO zero quark mixing and TBM neutrino mixing 
\cite{a4,Volkas}. The stability of the above vacuum alignment is
discussed in \cite{A4Warped}. The VEV of $\Phi$ induces an A$_4\to
Z_3$ SSB pattern, which in turn induces no quark mixing and is
peaked towards the IR brane. Similarly, the VEV of $\chi$ induces
an A$_4\to Z_2$ SSB pattern peaked towards the UV brane.
Subsequently, NLO interactions, including both $\Phi$ and $\chi$, break A$_4$ completely and induce
quark mixing and deviations from TBM. The Higgs VEV is in charge of the SSB pattern
$SU(2)_L\times SU(2)_R\rightarrow SU(2)_D$, which is peaked
towards the IR brane. The (gauge) SSB pattern on the UV brane is
driven by orbifold BC and a Planckian UV localized VEV, which is
effectively decoupled from the model \cite{Agashe:2003zs}.

\noindent To summarize the implications of the NLO interactions in
the quark and charged lepton sectors, we provide the structure of the LO+NLO 
mass matrices in the ZMA \cite{A4Warped}:
\begin{equation} \frac{1}{v}(M+\Delta M)_{u,d,\ell}^{4D}=\underbrace{\left(\begin{array}{ccc}y_{u,d,e}^{4D}&
 y_{c,s,\mu}^{4D} & y_{t,b,\tau}^{4D}
\\y_{u,d,e}^{4D}& \omega y_{c,s,\mu}^{4D} & \omega^2y_{t,b,\tau}^{4D}\\y_{u,d,e}^{4D}& \omega^2y_{c,s,\mu}^{4D} & \omega
y_{t,b,\tau}^{4D}\end{array}\right)}_{\sqrt{3}U(\omega)
diag(y_{u_i,d_i,e_i}^{4D})}+\left(\begin{array}{ccc}
f_\chi^{u,d,e}x_1^{u,d,\ell}&
f_\chi^{c,s,\mu}x_2^{u,d,\ell}&
f_\chi^{t,b,\tau}x_3^{u,d,\ell}\\0&0&0\\f_\chi^{u,d,e}y_1^{u,d,\ell}&
f_\chi^{c,s,\mu}y_2^{u,d,\ell}&
f_\chi^{t,b,\tau}y_3^{u,d,\ell}\end{array}\right),\label{MDeltaM}
\end{equation}
where $\omega=e^{2\pi i/3}$, $v=174$\,GeV is the 4D Higgs VEV,
$y^{4D}_{u,c,t,d,s,b,e,\mu,\tau}$ are the effective 4D LO  Yukawa
couplings, $(x_{i}^{\ell},y_i^\ell)\equiv (\tilde{x}_i^\ell,\tilde{y}_i^\ell)y_{u_i,d_i,e_i}^{4D}/y_{u_i,d_i,e_i}$ and $y_{u_i,d_i,e_i}$\,$(\tilde{x}_i^{u,d,\ell},\tilde{y}_i^{u,d,\ell})$
are the  5D LO\,(NLO) Yukawa couplings. The
function $f_\chi^{u_i,d_i,e_i}\simeq 2\beta_\chi
C_\chi/(12-c_{q_L,\ell_L}-c_{u_i,d_i,e_i})\simeq0.05$ accounts for the
suppression of the  NLO Yukawa
interactions and $C_\chi\!\equiv\!\chi_0/M_{Pl}^{3/2}\!\simeq\!0.155$. Finally,
the unitary matrix, $U(\omega)$ is the LO left diagonalization
matrix in both the up and down sectors, $(V_L^{u,d,\ell})_{LO}$,
which is independent of the LO Yukawa couplings, while
$(V_R^{u,d,\ell})_{LO}=\mathbbm{1}$(see \cite{A4Warped}). Using
standard perturbative techniques on the matrix in
Eq.~(\ref{MDeltaM}) we obtained $(V_{L,R}^{u,d,\ell})_{NLO}$
\cite{A4Warped,A4CPV} at
$\mathcal{O}\Big(f_\chi^{u_i,d_i,e_i}(\tilde{x}_i^{u,d,\ell},\tilde{y}_i^{u,d,\ell})\Big)$.
The left-handed diagonalization matrix is given by %%
\begin{equation}V_L^{q,\ell}=U(\omega)\left( \begin{array}{ccc} 1 &
f_\chi^{c,s,\mu}(\tilde{x}_2^{q,\ell}+\tilde{y}_2^{q,\ell}) &
f_\chi^{t,b,\tau}(\tilde{x}_3^{q,\ell}+\tilde{y}_3^{q,\ell})\\
-f_\chi^{c,s,\mu}[(\tilde{x}_2^{q,\ell})^* + (\tilde{y}_2^{q,\ell})^*] & 1 &
f_\chi^{t,b,\tau}(\tilde{x}_3^{q,\ell} + \omega \tilde{y}_3^{q,\ell})\\
-f_\chi^{t,b,\tau}[(\tilde{x}_3^{q,\ell})^* + (\tilde{y}_3^{q,\ell})^*]&
-f_\chi^{t,b,\tau}[(\tilde{x}_3^{q,\ell})^* + \omega^2(\tilde{y}_3^{q,\ell})^*]&
1\end{array}\right)\label{C3:VLQ},
\end{equation}
\noindent where $q=u,d$.

Similarly, the right diagonalization matrices  in the quark and charged lepton sectors, to first order in $f_\chi^{u_i,d_i,e_i} (\tilde{x}_i^{u,d,\ell},
\tilde{y}_i^{u,d,\ell})$ are given by %%
\begin{equation}V_R^{q,\ell}=\left( \begin{array}{ccc} 1 &
\Delta_1^{q,\ell} &
\Delta_2^{q,\ell}\\
-(\Delta_1^{q,\ell})^*  & 1 &
\Delta_3^{q,\ell}\\
-(\Delta_2^{q,\ell})^* & -(\Delta_3^{q,\ell})^* & 1
\end{array}\right)\label{VRQ},
\end{equation}
where $q=u,d$ and the $\Delta_i^{q,\ell}$ are  given by:
\begin{equation}
\Delta_1^{q,\ell}
=\frac{m_{u,d,e}}{m_{c,s,\mu}}\left[f_\chi^{u,d,e}\left((\tilde{x}_1^{q,\ell})^*
+\omega^2 (\tilde{y}_1^{q,\ell})^*\right)
+f_\chi^{c,s,\mu}\left(\tilde{x}_2^{q,\ell}+\tilde{y}_2^{q,\ell}\right)\right]
\label{DeltaR1},\end{equation}
\begin{equation}
\Delta_2^{q,\ell}
=\frac{m_{u,d,e}}{m_{t,b,\tau}}\left[f_\chi^{u,d,e}\left((\tilde{x}_1^{q,\ell})^*
+\omega (\tilde{y}_1^{q,\ell})^*
\right)+f_\chi^{t,b,\tau}\left(\tilde{x}_3^{q,\ell}+\tilde{y}_3^{q,\ell}\right)\right]
\label{DeltaR2},\end{equation}
\begin{equation}
\Delta_3^{q,\ell}
=\frac{m_{c,s,\mu}}{m_{t,b,\tau}}\left[f_\chi^{c,s,\mu}\left((\tilde{x}_2^{q,\ell})^*
+\omega (\tilde{y}_2^{q,e\ell})^*\right)
+f_\chi^{t,b,\tau}\left(\tilde{x}_3^{q,\ell}+\omega\tilde{y}_3^{q,\ell}\right)\right]
\label{DeltaR3}.\end{equation} %%
The suppression by quark mass ratios of the off-diagonal elements
in $V_R^{u,d}$, stemming from the degeneracy of LH bulk masses,  turned out to play an important role in relaxing
the flavor violation bounds on the KK mass scale, as compared to
flavor anarchic frameworks \cite{A4CPV}.

\section{The neutrino sector\,--\,Higher order corrections to the PMNS matrix and
$\theta_{13}$}\label{Sec:NUanalyze}

The global fits based on the recent measurements of $\nu_\mu\rightarrow\nu_e$
appearance in the RENO, Daya Bay, T2K, MINOS and other
experiments, allow one to obtain a significance of $10\sigma$ for
$\theta_{13}>0$, with best fit points at around
$\theta_{13}\simeq0.15$, depending on the precise treatment of
reactor fluxes  \cite{Fogli,Tortola}. We wish the RS-$A_4$ higher
order corrections to the PMNS matrix to be such that the new fits
are still ``accessible" by a significant portion of the model
parameter space.

In the RS-A$_4$ model deviations from TBM neutrino mixing are generated by higher order effects on the UV/IR branes and by ``cross-talk" and ``cross-brane" effects. In \cite{A4Warped} we studied in detail the textures induced by the three types of higher order corrections. Here we quote the  dominant corrections to the Dirac and Majorana mass matrices and the resulting modification of $V_L^\nu$ the left diagonalization matrix of the effective (LL) neutrino Majorana mass matrix.

Starting from the Dirac mass terms we have:
\beq
\mathcal{L}_\nu^{D}\subset k^{-1/2}y_\nu\bar{\ell}_LH\nu_R +k^{-2}y_\chi^D\bar{\ell}_LH\chi\nu_R
+k^{-7/2}y_D^{\Phi^2}\bar{\ell}_LH\Phi^2\nu_R, \label{DiracLagHO}\eeq

\noindent where $y_\nu$, $y_D^\chi$ and $y_D^{\Phi^2}$ are dimensionless 5D Yukawa couplings. The  4D effective 
mass matrix is obtained by integrating over $y$ the profiles of the fields/VEV's in each of the above operators and acquires the following form \cite{A4Warped}:
\beq \hat{M}_\nu^D=m_\nu^D\left(\!\!\begin{array}{ccc} 1+\epsilon_1 &\epsilon_2 &\epsilon_3+\epsilon_\chi\\\epsilon_3&1+
\epsilon_1
&\epsilon_2\\\epsilon_2+\epsilon_\chi&\epsilon_3&1+\epsilon_1\end{array}\!\!\right),\label{DiracMassMatrix}\eeq
where the $m_\nu^D$, $\epsilon_\chi$ and $\epsilon_{1,2,3}$ entries come from the first, second and third terms of Eq.~(\ref{DiracLagHO}), respectively. Notice that the smallness of $\epsilon_{\chi,1,2,3}$ compared to $m_\nu^D$ comes
not only from  the (mild) suppression of the flavon VEV's with respect to the 5D scale $k\sim\mathcal{O}(M_{Pl})$ but also
from the associated overlap correction factors \cite{A4Warped,A4CPV}. 

Turning to the Majorana mass terms we have:
\beq
\mathcal{L}_\nu^{M}\subset M\bar{\nu}_R^c\nu_R+ k^{-1/2}y_\chi\chi \bar{\nu}_R^c\nu_R+k^{-2}y_{\chi^2}\chi^2\bar{\nu}_R^c\nu_R,
\label{MajoLagHO}\eeq

The resulting 4D (heavy) Majorana mass matrix acquires the following form:
\beq \hat{M}_\nu^M=\tilde{M}\left(\!\!\begin{array}{ccc} 1+\epsilon_4 & 0 & M_\chi/\tilde{M}\\ 0&1+\epsilon_5&0\\
M_\chi/\tilde{M}&0&1+\epsilon_4^*\end{array}\!\!\right),\label{MajoMassMatrix}\eeq

\noindent where  $\tilde{M}$, $M_\chi$ and $\epsilon_{4,5}$ come from the first, second and third terms of Eq.~(\ref{MajoLagHO}).
The smallness of $\epsilon_{4,5}$ comes mainly from the suppression associated with the flavon VEV's, since the effect of 
overlap corrections is milder for UV peaked operators \cite{A4Warped}.

The effective (light) Majorana mass matrix is now obtained using the see-saw  mechanism, $\hat{M}_\nu^{eff}=-(\hat{M}_\nu^D)^T(\hat{M}_\nu^M)^{-1}\hat{M}_\nu^D$. If we turn off $\epsilon_{1-5,\chi}$ we obtain the LO results of \cite{A4Warped}, for which the mass spectrum is given by \beq (\hat{M}_{eff.}^\nu)_{diag.}=(V_L^\nu)^T\hat{M}_{eff.}^\nu V_L^\nu=-\frac{(m_\nu^D)^2}{\tilde{M}}{\rm diag}\left[\frac{1}{1+q},1,\frac{1}{1-q}\right] \label{NumassSpectrum}\eeq

\noindent where $q\equiv M_\chi/\tilde{M}$. Notice that the overall neutrino mass scale is set by the ratio $(m_\nu^D)^2/\tilde{M}$, while the type of hierarchy and its extent are determined by $q$. Using $\Delta m_{atm}^2\simeq 2.43\cdot10^{-3}{\rm eV}^2$ and
$\Delta m_{sol}^2\simeq 7.6\cdot10^{-5}{\rm eV}^2$ from \cite{Fogli}, we are able to obtain a cubic equation for $q$ with solutions
$q\simeq[-2.02, -1.98, 0.79, 1.19]$, where the first (last) two solutions correspond to inverted (normal) mass hierarchy \cite{A4Warped}. The LO diagonalization matrix $V_L^{\nu}$ is given by:
\beq V_L^{\nu}=\left(\begin{array}{ccc} 1/\sqrt{2} & 0 & -1/\sqrt{2} \\0&1&0\\1/\sqrt{2}&0& 1/\sqrt{2}\end{array}\!\!\right).\eeq

We now proceed to obtain the corrections to $V_L^\nu$ considering the higher order effects encoded in Eqs. (\ref{DiracMassMatrix})
and (\ref{MajoMassMatrix}). Since numerically $\epsilon_{1-5,\chi}\sim 0.05$, we can perform (analytically) the perturbative diagonalization
of $(\hat{M}_\nu^{eff.})_{NLO}$ to leading order in each of the $\epsilon$'s. The resulting $(V_L^\nu)_{NLO}\equiv\tilde{V}_L^\nu$ is given by:
\beq \tilde{V}_L^\nu=\left(\!\!\begin{array}{ccc} \frac{1}{\sqrt{2}}+\frac{\epsilon_2^*-\epsilon_3^*}{2\sqrt{2}}-i\frac{q^2-1}{2\sqrt{2}q}{\rm Im}(\epsilon_4) & 
\frac{\epsilon_2+\epsilon_3}{q}+i\,{\rm Im}(\epsilon_2)+{\rm Re}(\epsilon_3)& -\frac{1}{\sqrt{2}}+\frac{\epsilon_2-\epsilon_3}{2\sqrt{2}}+i\frac{q^2-1}{2\sqrt{2}q}{\rm Im}(\epsilon_4)\\
-\frac{2+q}{\sqrt{2}q}\left(\epsilon_2^*+\epsilon_3^*\right)\phantom{aaa} & 1\phantom{/} & \frac{\epsilon_3-\epsilon_2}{\sqrt{2}}\\
\frac{1}{\sqrt{2}}-\frac{\epsilon_2^*-\epsilon_3^*}{2\sqrt{2}}+i\frac{q^2-1}{2\sqrt{2}q}{\rm Im}(\epsilon_4) & \frac{\epsilon_2+\epsilon_3}{q}+i\,{\rm Im}(\epsilon_3)+{\rm Re}(\epsilon_2) &\frac{1}{\sqrt{2}}+\frac{\epsilon_2-\epsilon_3}{2\sqrt{2}}+i\frac{q^2-1}{2\sqrt{2}q}{\rm Im}(\epsilon_4)\end{array}\!\!\right). \label{VLnuNLO}\eeq

Using $V_L^\ell$ of Eq.~(\ref{C3:VLQ}) and $(\tilde{V}_L^\nu)$ of Eq.~(\ref{VLnuNLO}), we obtain the modified PMNS matrix, which includes all dominant higher order effects from the charged lepton and neutrino sectors, $V_{PMNS}\equiv W=(V_L^{\ell})^\dagger \tilde{V}_L^\nu$. Notice  that all elements of the PMNS matrix will be still independent of the LO
Yukawa couplings $(y_{e,\mu,\tau,\nu,\chi,M})$, due to the Form diagonalizability of the LO mass  matrices. We avoid writing the resulting PMNS matrix explicitly, due to its complexity. By expanding the basis independent relations \beq \theta_{13}={\rm arcsin}|W_{13}|, \quad \theta_{12}={\rm arctan}(|W_{12}|/|W_{11}|),\quad
 \theta_{23}={\rm arcsin}(|W_{23}|/|W_{33}|),\label{PMNSrelations}\eeq  we are able to obtain approximate analytical expressions for  $\theta_{12,13,23}$,

\beq \sin \theta_{12}\simeq\frac{1}{\sqrt{3}}\left|1+\frac{2+q}{3q}\left(\phantom{\frac{1}{2}}\!\!\!\!\epsilon_2+\epsilon_3+2{\rm Re}(\epsilon_2+\epsilon_3)\right)-\tilde{f}_\chi^{\mu,\tau}\left(\phantom{\frac{1}{2}}\!\!\!\!\omega^2
(x_2^\ell+y_2^\ell)+\omega(x_3^\ell+y_3^\ell)\right)\right|\label{T12},\eeq

\beq \sin\theta_{13}\simeq\frac{1}{\sqrt{6}}\left|\tilde{f}_\chi^{\mu,\tau}\left[\phantom{\frac{1}{2}}\!\!\!\!(x_2^\ell+y_2^\ell)(1-\omega)+(x_3^\ell+y_3^\ell)(1-\omega^2)\right]+\frac{i(q^2-1)}{q}{\rm Im}\,\epsilon_4\right|\label{T13},\eeq

\beq \sin\theta_{23}\simeq\frac{1}{\sqrt{2}}\left|-\omega+i\sqrt{3}(\epsilon_2-\epsilon_3)+\tilde{f}_\chi^{\mu,\tau}
\left(\phantom{\frac{1}{2}}\!\!\!\!\omega^2x_3^\ell+\omega x_3^{\ell *}+2{\rm Re}(y_3^\ell)\right)+\frac{i(q^2-1)}{\sqrt{3}q}{\rm Im}\,\epsilon_4\right|.\label{T23}\eeq

\begin{center}
\begin{figure}
\includegraphics[width=7.88cm]{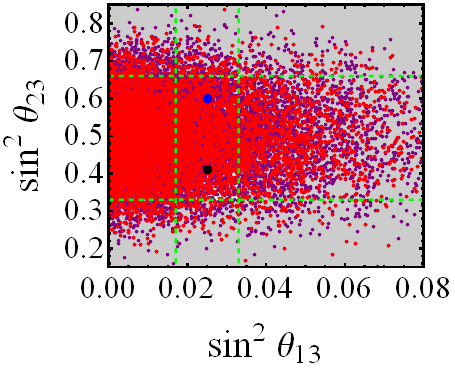}\quad\quad\!\!\!\!
\includegraphics[width=7.888cm]{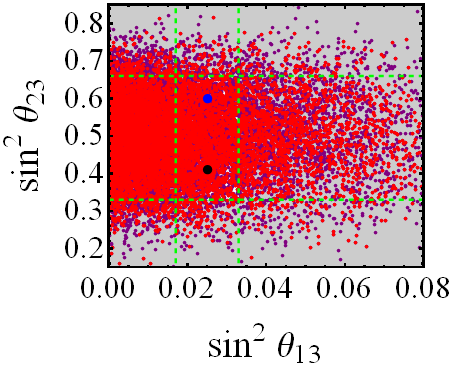}\caption{\it
Model predictions for $\theta_{23}$ vs. 
$\theta_{13}$ for normal (left) and inverted (right) mass hierarchy, including all dominant higher order and
cross talk effects. The purple points represent the entire sample, while the red points satisfy the $3\sigma$ bounds on $\theta_{12}$. The  rectangles, confined between the horizontal and vertical dashed green lines, represent the $3\sigma$
allowed regions from the global fits of
\protect\cite{Fogli,Tortola}.}\label{NUAngles1}
\end{figure}
\end{center}

\begin{center}
\begin{figure}
\includegraphics[width=7.88cm]{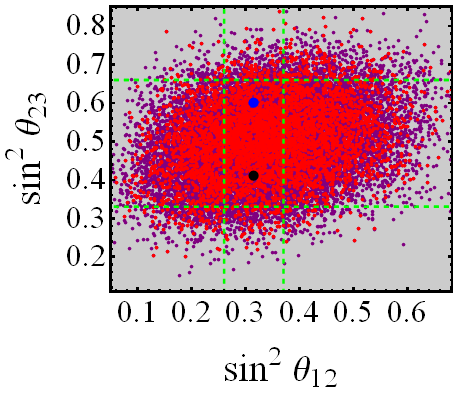}\quad\quad\!\!\!\!
\includegraphics[width=7.88cm]{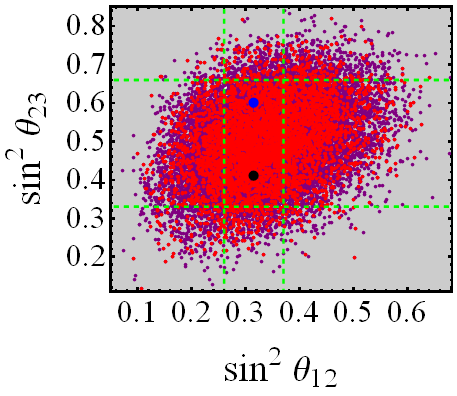}\caption{\it
Model predictions for $\theta_{23}$ vs. 
$\theta_{12}$ for normal (left) and inverted (right) mass hierarchy, including all dominant higher order and
cross talk effects. The purple points represent the entire sample, while the red points satisfy the $3\sigma$ bounds on $\theta_{13}$. The  rectangles, confined between the horizontal and vertical dashed green lines, represent the $3\sigma$
allowed regions from the global fits of
\protect\cite{Fogli,Tortola}.}\label{NUAngles3}
\end{figure}
\end{center}

We are now ready to perform a systematic study of the RS-A$_4$ predictions for $\theta_{12,13,23}$.
To this end we recall that $\epsilon_{1-5,\chi}$ of Eqs.~(\ref{DiracMassMatrix}) and (\ref{MajoMassMatrix}),
implicitly contain the dimensionless Yukawa couplings of their corresponding operator in Eqs. (\ref{DiracLagHO}) and (\ref{MajoLagHO}). For this reason $\epsilon_{1-5,\chi}$ will be multiplied by an $\mathcal{O}(1)$ complex Yukawa coupling for the purpose of the numerical scan below. We generate  sample of 20000 points, in which all Yukawas $(\tilde{x}_i^\ell,\tilde{y}_i^\ell,y_{\epsilon_i})$ are complex numbers with random phases and magnitudes normally distributed around $1$ with standard deviation, $\sigma=0.5$.
Using the overlap integrals performed in \cite{A4Warped,A4CPV}, we get $\epsilon_\chi\simeq0.07$, $\epsilon_{1,2,3}\simeq0.04, \epsilon_{4,5}\simeq0.09$ and $\tilde{f}_\chi^{e,\mu,\tau}\equiv\sqrt{3}f_\chi^{e,\mu,\tau}\simeq0.08$. By Substituting $\epsilon_{1-5,\chi}$ in  Eq.~(\ref{PMNSrelations}), we study the correlations between the various neutrino mixing angles. The results of the scan are plotted in Figs. \ref{NUAngles1}\,--\ref{NUAngles3}. The rectangle
confined between the dashed green lines represent the $3\sigma$ allowed regions from the global fits of \cite{Fogli,Tortola}.
The purple points in each plot represent the entire sample, while the red points, depicted on top of the purple ones, satisfy
in addition the $3\sigma$ constraint from the mixing angle not appearing in the same figure. In other words, the red points in Fig.~
\ref{NUAngles1} satisfy the $3\sigma$ constraint on $\theta_{12}$ and so on. The best fit point(s) are depicted in black (blue) for
$(\theta_{23})_{{\rm best}}$ values from the first (second) octant, where $\Delta\chi^2_{{\rm black}}<\Delta\chi^2_{{\rm blue}}$ in the global fit of \cite{Fogli}.

We realize that there are only mild differences between the NH and IH cases, coming from the terms proportional to
$(2+q)$ and $(q^2-1)$ in Eqs.~(\ref{T12})\,--\,(\ref{T23}). Qualitatively, this means that the deviations from TBM mixing in the IH case are slightly larger  for $\theta_{13,23}$ and slightly smaller for $\theta_{12}$, as can be seen in Figs.\,\,\ref{NUAngles1}\,--\ref{NUAngles3}. To quantify the viability of the RS-A$_4$ setup in predicting realistic neutrino mixing angles, we can define a success rate for the NH and IH cases in analogy with \cite{FeruglioNew}. Namely, we ask ourselves what is the percentage of the points, $\xi$, satisfying the $3\sigma$ constraints for all neutrino mixing angles. We get $\xi_{NH}\simeq10\%$ and $\xi_{IH}\simeq12.5\%$, which is slightly better  than the results for typical A$_4$  models in \cite{FeruglioNew}.
We avoid analyzing the RS-A$_4$ predictions for $\delta_{CP}^\nu$, due to the lack of sufficient experimental data and its potential sensitivity to higher order effects of $\mathcal{O}(\epsilon_i^2,f_\chi^2)$. The above analysis re-demonstrates \cite{Flasy2011,Flasy2012} the viability of models predicting TBM at LO, with corrections coming from higher order effects in both the charged lepton and neutrino sectors  and in particular the RS-A$_4$ setup.

\begin{center}
\begin{figure}
\includegraphics[width=7.88cm]{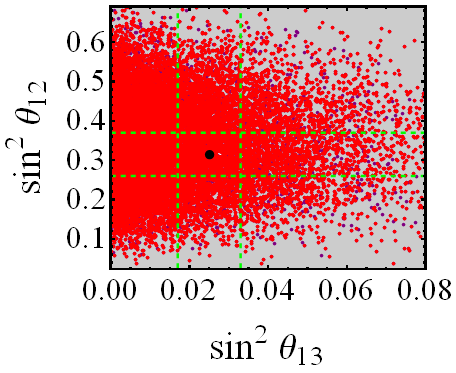}\quad\quad\!\!\!\!
\includegraphics[width=7.88cm]{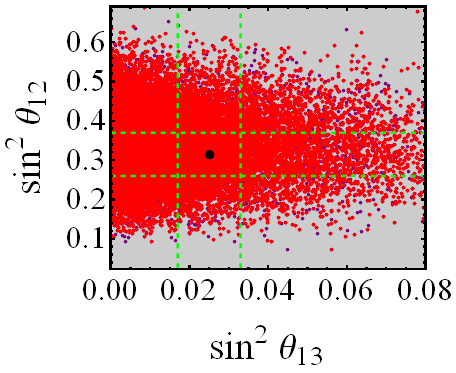}\caption{\it
Model predictions for $\theta_{12}$ vs. 
$\theta_{13}$ for normal (left) and inverted (right) mass hierarchy, including all dominant higher order and
cross talk effects. The purple points represent the entire sample, while the red points satisfy the $3\sigma$ bounds on $\theta_{23}$. The rectangles, confined between the horizontal and vertical dashed green lines, represent the $3\sigma$
allowed regions from the global fits of
\protect\cite{Fogli,Tortola}.}\label{NUAngles2}
\end{figure}
\end{center}

\subsection{Simplifications in the brane localized RS-A$_4$ setup}\label{Sec:Nubrane}
If we specialize to the brane localized version of the RS-A$_4$ model \cite{Csaki:2008qq,JoseA4} the cross talk interactions associated with the $\tilde{x}_{2,3}^{\ell}$ and $\tilde{y}^{\ell}_{2,3}$ are absent and the expressions for the mixing angles simplify to the following form:
\beq
 \sin\theta_{13}\simeq\left|\frac{i(q^2-1)}{\sqrt{6}q}{\rm Im}\,\epsilon_4\right|,\qquad \sin \theta_{12}\simeq\left|1+\frac{2+q}{3\sqrt{3}q}\left(\phantom{\frac{1}{2}}\!\!\!\!\epsilon_2+\epsilon_3+2{\rm Re}(\epsilon_2+\epsilon_3)\right)\right|\label{NuAnglesLO1312}\eeq
\beq \sin\theta_{23}\simeq\left|\frac{1}{2\sqrt{2}}+i\left(-\frac{\sqrt{3}}{2\sqrt{2}}+\sqrt{\frac{3}{2}}(\epsilon_2-\epsilon_3)+\frac{(q^2-1)}{\sqrt{6}q}{\rm Im}\,\epsilon_4\right)\right|\label{NuAnglesLO23}.\eeq

\noindent We immediately realize that $\theta_{13}$ is now governed by $q$ and ${\rm Im}(\epsilon_4)$. We take $\epsilon_4$ to be pure imaginary and fix its magnitude to obtain the best fit value $\sin^2(\theta_{13})\simeq0.158$ \cite{Fogli,Tortola}. The fixing of $\epsilon_4$ depends on the type of  hierarchy, as determined by  $q=M_\chi/\tilde{M}$ (Eq.~(\ref{NumassSpectrum})). For the NH case we get ${\rm Im}(\epsilon_4)\simeq0.5 (0.97)$ for $q\simeq0.79\,(1.19)$, while for the IH case we have ${\rm Im}(\epsilon_4)\simeq 0.6$ for $q=-2.02\,(-1.98)$. These solutions for ${\rm Im}(\epsilon_4)$ implicitly include $y_{\epsilon_4}$ , the dimensionless 5D Yukawa coupling of the operator $k^{-2}\chi^2\bar{\nu}_R^c\nu_R$, entering as $\epsilon_4\equiv k^{-1} y_{\epsilon_4}v_\chi^{UV}$ in the 4D effective theory on the UV brane. For the characteristic value $k^{-1}v_\chi^{UV}\simeq 0.1$ we have $|y_{\epsilon_4}^{NH}|\simeq 5\,(10.5)$ and $|y_{\epsilon_4}^{IH}|\simeq 6$, all of which satisfy the naive dimensional analysis (NDA) perturbativity bounds from \cite{Csaki:2008qq} $|y_{\epsilon_4}|<4\pi$.

 Now that we solved for $\theta_{13}$, the values of $\theta_{12,23}$ are controlled by the $\epsilon_{2,3}$ parameters coming from the operator $k^{-2}\bar{\ell}_LH\Phi^2\nu_R$. In general, we want $\theta_{23}$ to deviate towards its best fit value in the first octant $\sin^2\theta_{23}^{bestI}\simeq 0.427$ \cite{Tortola} and $\theta_{12}$ stay close to its trimaximal value. Observing Eq.~(\ref{NuAnglesLO23}), we realize that $\theta_{13}$ and $\theta_{23}$ contain exactly the same correction term proportional to ${\rm Im}(\epsilon_4)$,  fixed by the value of $\theta_{13}$. Thus, we have a strong correlation between the increase of $\theta_{13}$ from zero and the deviation of $\theta_{23}$ towards the first (or second) octant. Since $\epsilon_{2,3}$ are of the same strength, we can  take them to be pure imaginary and opposite in sign to change $\theta_{23}$ to the desired value $(\theta_{23}^{bestI})$ while keeping $\sin\theta_{12}\simeq1/\sqrt{3}$. Since the latter  possibility is highly fine tuned, we find it more instructive to consider the more general case in which $\epsilon_{2,3}\simeq0.04\,y_{\epsilon_{2,3}}$, where $y_{\epsilon_{2,3}}$ are complex numbers with random phases and magnitudes normally distributed around 1 with standard deviation of 0.5. In this case, $\theta_{12,23}$ are governed by four parameters, two phases and two Yukawas. We first specify to the case in which the magnitudes  are fixed, where the number of observables equals the number of input parameters, in order to look for a pattern (see also \cite{MuChun2012}).

\begin{center}
\begin{figure}
\includegraphics[width=7.25cm]{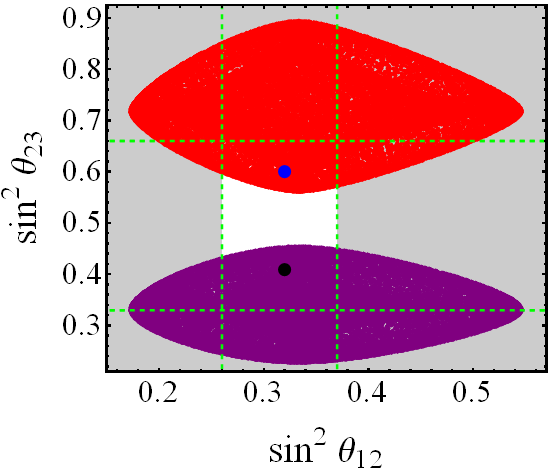}\quad\quad\!\!\!\!
\includegraphics[width=7.88cm]{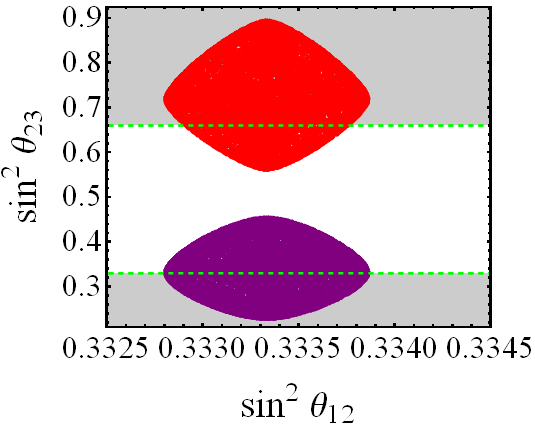}\caption{\it
The brane localized RS-A$_4$ predictions  for $\theta_{12}$ vs. 
$\theta_{23}$ for normal (left) and inverted (right) mass hierarchy, including all dominant higher order
effects. The best fit value $\sin(\theta_{13})\simeq 0.158$  fixes ${\rm Im }(\epsilon_4)$ up to a sign. The purple (red) points correspond to a positive (negative) value of ${\rm Im}(\epsilon_4)$ and we scan over the phases of $\epsilon_{2,3}$ fixing their magnitudes to 0.04 to make the comparison with the ``cross-talk" case more transparent. The white rectangles, confined between the horizontal and vertical dashed green lines, represent the $3\sigma$
allowed regions from the global fits of
\protect\cite{Fogli,Tortola}.}\label{NUAngles32}
\end{figure}
\end{center}

\begin{center}
\begin{figure}
\includegraphics[width=7.35cm]{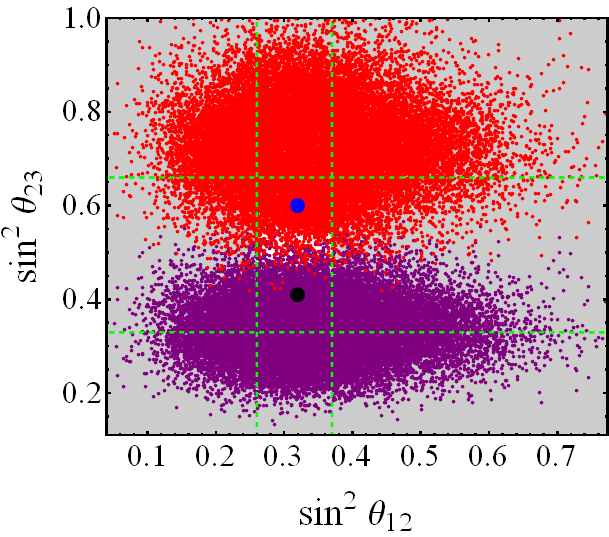}\quad\quad\!\!\!\!
\includegraphics[width=7.88cm]{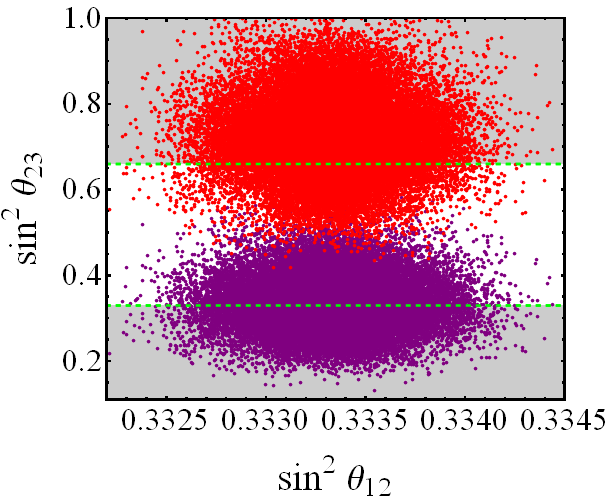}\caption{\it
The brane localized RS-A$_4$ predictions  for $\theta_{12}$ vs. 
$\theta_{23}$ for normal (left) and inverted (right) mass hierarchy, scanning over the phases and magnitudes of $\epsilon_{2,3}$. The best fit value $\sin(\theta_{13})\simeq 0.158$  fixes ${\rm Im }(\epsilon_4)$ up to a sign. The purple (red) points correspond to  positive (negative) values of ${\rm Im}(\epsilon_4)$. The white rectangles, confined between the horizontal and vertical dashed green lines, represent the $3\sigma$
allowed regions from the global fit of
\protect\cite{Fogli,Tortola}.}\label{NULOscan}
\end{figure}
\end{center}

\noindent We depict the results in Fig.~\ref{NUAngles32}. It can be seen that $\theta_{23}$ moves strongly to the first  (second) octant if ${\rm Im}(\epsilon_4)$ is positive (negative) and that the best fit point is rather far from the center of the resulting distributions. It is not sensible to talk about a success rate in such a case due to the assignment imposed by $\theta_{13}$ and the simplifying assumptions on $\epsilon_{2,3}$. Relaxing the assumption on the magnitudes of $\epsilon_{2,3}$  have the effect of spoiling the tendency towards the first or second octant, as can be seen by the slight overlap
between the ``positive" and ``negative" cases, as can be seen in Fig.~\ref{NULOscan}.

 The analysis of this section shows that the brane localized version of RS-A$_4$ is
more constrained by the experimental neutrino data, due to the fact that the Yukawa coupling associated with $\epsilon_4$ is fixed by $\theta_{13}$ to values of order $|y_{\epsilon_4}|\approx 6$ which are still below the perturbativity bound implied by NDA but still render the 5D theory itself less accurate. On the other hand, the small number of parameters $(\epsilon_{2,3,4})$ implies significant simplifications and an interesting correlation between the value of $\theta_{13}$ and  the (bi-directional) deviations from maximality of $\theta_{23}$. This feature is attributed to the A$_4$ assignments and vacuum alignment. Finally, the corrections to $\theta_{12}$ in the IH case are extremely small due to the absence of $\epsilon_4$ and the $(2+q)/q$ term in Eq.~(\ref{T12}).

\section{The charged lepton sector- Anomalous Z couplings and cLFV}\label{Sec:Zcouplings}

Charged lepton flavor violating processes like $\mu\to 3e$ and $\mu\to e$ conversion in the presence of a nucleus
are the main focus of a series of  experiments,  planned to be performed in J-PARC \cite{DeeMee,COMET,PRIME}, Osaka (RNCP) \cite{MuSIC}, Fermilab \cite{Mu2e,ProX} and PSI \cite{MU3e}, within the next decade and increase the accuracy for these searches by roughly five orders of magnitude. The current upper bounds were set by the SINDRUM $(\mu\to 3e)$  \cite{sindrum} and SINDRUM II 
$\left(\mu\to e \,\,{\rm in} \,\,{\rm Ti}_{22}^{48}\right)$\cite{sindrumII} experiments.

As stated in the introduction the main source of charged lepton
flavor violation (cLFV) in RS-$A_4$ are anomalous Z couplings,
induced by gauge and fermionic KK mixing, generating Tree level Z exchange
contributions to $\mu\to e,3e$, $B_s\to\mu^+\mu^-$ and other processes. The
effect of EWSB on the mixing of the Z boson with its KK
partners and those of the (custodial) $Z^\prime$, can be studied
directly in the zero mode approximation (ZMA), by solving the $Z$ equations of motion in the vicinity of an IR boundary term which induces the SSB pattern $SU(2)_L\times SU(2)_R\to SU(2)_D$ (e.g. \cite{CsakiZold}).
On the other hand, the effect of KK fermion mixing  has to be studied
by considering  the full KK mass matrices \cite{BurasKK}.

\subsection{The impact of gauge boson mixing on Z couplings}

The effect of EWSB on gauge bosons mixing translates via the EOM to a
distortion of the $Z$ wave function near the IR brane . The advantage in using this procedure is that 
all gauge bosons and their KK partners are automatically given in the physical basis. The canonically normalized 
physical Z boson  wave function (profile) is then given by \cite{CsakiZold}

\begin{equation}
f_Z^{(0)}(y) \simeq \frac{1}{\sqrt{\pi R}}\left[\phantom{\frac{1}{2}}\!\!\!\!1+ \frac{M_Z^2e^{2ky}}{4k^2}\left(\phantom{\frac{1}{2}}\!\!\!\!1-2 ky \right)\right].\label{Zprofile}
\end{equation}

The modified $Z$ wave function will induce small deviations in the $Z$ coupling matrices $A_{L,R}(Z)$, due 
to the different overlaps of the zero and KK modes profiles with $f_z^{(0)}(y)$. Schematically the overlap integrals  associated with the $Z$ couplings are written as:
\bea 
\frac{(g_{L,R}^Z)}{\sqrt{\pi R}}&\!\!\!=\!\!\!&(g_{L,R}^Z)_{SM}R_{nm}^Z\nonumber\\&\!\!\!=\!\!\!&(g_{L,R}^Z)_{SM}\int dy f^{(0)}_Z(y)\hat{\chi}_{L,R}^{(0,1,1^{--},1^{+-},1^{-+}\,...)}(c_{\ell,(e,\mu,\tau)},y)
\hat{\chi}_{L,R}^{(0,1,1^{--},1^{+-},1^{-+}\,...)}(c_{\ell,(e,\mu,\tau)},y),\nonumber\\
\label{Zintegral}\eea
where $(g_{L,R}^Z)_{SM}=(T^3_L)_{\ell,e,\mu,\tau}-Q_{em}^2\sin^2\theta_W$ and $\hat{\chi}_{L,R}^{(m)}(c_{\ell,e,\mu,\tau},y)$ are the canonically normalized wave functions for the zero mode fermions and their KK partners  (denoted by $m$ and $n$).
 
Before EWSB the $Z$ profile is flat and as a consequence the overlap integrals $R_{nm}^Z$ reduce to an integral over the $n^{th}$ and $m^{th}$ KK modes,  satisfying $R_{nm}^Z(f_Z^{(0)}=1/\sqrt{\pi R})=\delta_{nm}$ by virtue of the orthonormality of the  wave functions and implying $[A_{L,R}(Z)]\propto\mathbbm{1}$\footnote{At this stage we ignore the presence of opposite chirality $(--)$ KK fermions and $(+-,-+)$``custodians", which will be thoroughly discussed in the next section.}. When EWSB takes place, $f_Z^{(0)}$ is given by Eq.~(\ref{Zprofile}) and the  coupling matrices, $[A_{L,R}(Z)]$ are no longer proportional to the identity. As a result,  once the $Z$ coupling matrices will be rotated to the physical basis of the zero+KK fermions, new off diagonal entries will be generated in  $A_{L,R}(Z)$, corresponding to flavor and KK number (and BC type) violating $Z$ couplings.

We first focus on the gauge boson contribution to flavor violating Z couplings in the ZMA, since in this case we are able to obtain 
simple analytical expressions using the left and right diagonalization matrices of Eqs.~(\ref{C3:VLQ}) and (\ref{VRQ})\,--\,(\ref{DeltaR3}).
We start by performing the integral in Eq.~(\ref{Zintegral}) for the zero modes of $\ell_L$ and $e_R^{(\prime,\prime\prime)}$ (Eq.~(\ref{FermionContent}), which are identified with the charged lepton sector of  the SM. Using 
 \begin{equation}
\hat{\chi}^{(0)}_{L,R}(c,y)=\sqrt{\frac{k(1-2c)}{e^{k\pi R(1-2c)}-1}}e^{(1/2-c)k|y|},\quad \hat{\chi}^{(0)}_{L,R}(c,y=\pi R)\equiv \sqrt{2k}F(c),\label{zeromode} \eeq
the resulting deviations in the diagonal entries of the $Z$ coupling matrix are given by:
 \beq g^{Zff}=\sqrt{\pi R}\,(g_{L,R}^Z)_{SM}R_{00}^Z\simeq (g^{Z}_{L,R})_{\text{SM}}\left[1-\frac{M_Z ^2F^2(c) }{\Lambda_{IR}^2(3-2c)} \left(k\pi R-\frac{5-2c}{2(3-2c)}\right)\right],\label{GaugeDeviation}
\end{equation}
\noindent  where $f=\ell,e,\mu,\tau$ and $c=c_{\ell,e,\mu,\tau}$.  The physical couplings are obtained by rotating  the  $Z$ coupling  matrices for the zero modes, $\left[A_{L,R}(Z)\right]_{00}^G$ to the mass eigenbasis:
\bea \left(\left[A_L(Z)\right]_{00}^G\right)^{mass}\!\!\!&=&\!\!\!(V_L^\ell)^\dagger\left[A_L(Z)\right]_{00}^G V_L^\ell\equiv \sqrt{\pi R}
\,(g^Z_L)_{SM}(V_L^\ell)^\dagger \left(R_{00}^Z(c_\ell)\cdot\mathbbm{1}\right)V_L^\ell,\label{gLgauge}\\
\left(\left[A_R(Z)\right]_{00}^G\right)^{mass}\!\!\!&=&\!\!\!(V_R^\ell)^\dagger \left[A_R(Z)\right]_{00}^G V_R^\ell\equiv\sqrt{\pi R}\,(g^Z_R)_{SM}(V_R^\ell)^\dagger   {\rm diag}
[R_{00}^Z (c_{e,\mu,\tau})]V_R^\ell. \label{GaugeCorrMass}\eea

\noindent Since $\left[A_L(Z)\right]_{00}^G\propto\mathbbm{1}$, no LH flavor violating Z couplings will be generated through gauge boson 
mixing after EWSB and in particular $\delta (g_L^{\mu e})_{{\rm gauge}}=0$. On the other hand,  the non degeneracy of RH bulk masses imply $\left[A_R(Z)\right]_{00}^G\!
\propto\!\!\!\!\!\!\!/\,\,\mathbbm{1}$ and hence $(\delta g_R^{\mu e})_{{\rm gauge}}\neq0$. By using Eqs.~(\ref{VRQ})\,--\,(\ref{DeltaR1}) and (\ref{GaugeDeviation})\,--\,(\ref{GaugeCorrMass}) we are able to estimate the strength of this effect

\beq \delta g_R^{\mu e}\equiv (g_R^{Z\mu e})_{RS-A_4}-\underbrace{(g_R^{Z\mu e})_{SM}}_{\equiv 0}=\left[(V_R^\ell)^\dagger \left[A_R(Z)\right]_{00}^G V_R^\ell\right]_{12,21}\simeq \Delta_1^\ell\left(\delta g_R^{Z\mu\mu}-\delta g_R^{Zee}\right),\label{ZMuEgauge}
\eeq

\noindent where $\delta g_R^{Z\mu\mu,Zee}$ are defined in an analogous way and extracted directly from Eq.~(\ref{GaugeDeviation}).
Since $F(c_e)/F(c_\mu)\approx m_e/m_\mu$, the dominant contribution to $(\delta g_R^{\mu e})_{gauge}$ will come from $\delta g_R^{Z\mu\mu}$. We realize that $(\delta g_R^{\,u e})_{gauge}$ is suppressed by both $F^2(c_\mu)M_Z^2/\Lambda_{IR}^2$,  and $f_\chi^{\mu}(m_e/m_\mu) $. For the characteristic values, $\Lambda_{IR}\simeq 1.5\,  {\rm TeV}$, $c_\ell\simeq0.51$ and $c_\mu\approx0.63$, we have $\delta (g_R^{\mu e})_{gauge}\approx 5\times10^{-11}$.

\subsection{The impact of KK fermion mixing on Z couplings}

To account for  KK mixing effects, we write the LO KK mass matrix for the first
generation in the charged lepton sector in analogy with
\cite{IsidoriPLB,A4CPV}. This mass matrix includes the zero and first level
KK modes and acquires the following form: \cite{A4CPV}
\begin{equation}\frac{\hat{\mathbf{M}}_e^{KK}}{(M_{KK})}=
\left(\begin{array}{c}
\bar{\ell}_L^{e(0)} \\ \bar{\ell}^{e(1)}_L\\\bar{e}_L^{(1^{--})}\\
\bar{\tilde{e}}_L^{(1^{+-})} \end{array} \right)^T \!\!\left(
\begin{array}{cccc}
\breve{y}_eF_\ell F_e r_{00} x & 0 &\breve{y}_eF_\ell r_{01} x &
\breve{y}_\nu F_\ell r_{101} x \\
 \breve{y}_eF_e r_{10} x  &  1& \breve{y}_e^*r_{22} x  & \breve{y}_\nu r_{111} x \\
 0 &  \breve{y}_er_{11} x & 1 &  0 \\
 0 &  \breve{y}_\nu ^*r_{222} x & 0 & 1
\end{array}
\right)\left(\begin{array}{c}
e_R^{(0)}\\\ell_R^{e(1^{--})} \\ e^{(1)}_R\\
\tilde{e}_R^{(1^{-+})} \end{array} \right),\label{M4KK}
\end{equation}
where we factorized a common KK mass scale $M_{KK}\simeq2.45\Lambda_{IR}$,
$\breve{y}_{e,\nu}\equiv 2y_{e,\nu}v_\Phi^{4D}/\Lambda_{IR}$ and
the perturbative expansion parameter is defined as $x\equiv
v/M_{KK}$. In the above equation the various $r$'s
denote the ratio of the bulk and IR localized effective couplings
of the modes corresponding to the matrix element in question. For
simplicity, we define $r_{111}\equiv r_{11^{-+}}$, $r_{101}\equiv
r_{01^{-+}}$, $r_{22}\equiv r_{1^-1^-}$, $r_{222}\equiv
r_{1^{-}1^{+-}}$ and the notation for the rest of the overlaps is
straightforward \cite{A4CPV}. The associated Yukawa matrix,
$\hat{Y}^{e}_{KK}$,  is obtained by simply eliminating $x$ and the
1's from the above matrix and once rotated to the mass basis it describes the physical coupling between
$(++)$, $(--)$, $(+-)$ and $(-+)$ KK modes.  In \cite{A4CPV} we have performed an analytical diagonalization of all
one-generation zero+KK mass matrices in the quark sector and showed that the mild differences between the generations, stem from the  structure of the 5D Yukawa couplings and non degeneracy of the RH bulk mass parameters. 

The full three generation mass matrix will be $12\times 12$ and of
similar structure, which is modified mainly by the $A_4$ flavor
structure. We first decompose  the  $12\times12$ mass matrix of the RS-A${}_4$
charged lepton sector in terms of the one-generation KK+zero matrices in
Eq.~(\ref{M4KK})
\begin{equation}
\hat{\bf{M}}^{\ell}_{Full}=M_{KK}\left(\begin{array}{ccc}
\hat{\bf{M}}_e^{KK}/M_{KK} &
x\hat{Y}^{\mu}_{KK}(\hat{y}^{LO}_{12},F_\mu)
&x\hat{Y}^{\tau}_{KK}(\hat{y}^{LO}_{13},F_\tau)\\
x\hat{Y}^{e}_{KK}(\hat{y}^{LO}_{21},F_e) &
\hat{\bf{M}}_\mu^{KK}/M_{KK} &
x\hat{Y}^{\tau}_{KK}(\hat{y}^{LO}_{23},F_\tau)\\x\hat{Y}^{e}_{KK}(\hat{y}^{LO}_{31},F_e)
&x\hat{Y}^{\mu}_{KK}(\hat{y}^{LO}_{32},F_\mu)&\hat{\bf{M}}_\tau^{KK}/M_{KK}
\end{array}\right)\,,\label{MFullD}\end{equation}
where the expression in parenthesis for each off-diagonal element
denotes the replacements to be made in Eq.~(\ref{M4KK}) and the $\hat{\mathbf{M}}_{\mu,\tau}^{KK}$ matrices
are of exactly the same structure in Eq.~(\ref{M4KK}). 
 To account for NLO Yukawa interactions, the replacement $\hat{y}^{LO}_{ij}\rightarrow
\hat{y}^{LO}_{ij}+\hat{y}^{NLO}_{ij}$ for the dimensionless 5D Yukawa coupling matrices applies. In the above
equation, $M_{KK}$ is the KK mass corresponding to the degenerate
left-handed bulk mass parameter, $c_\ell^L$, and we normalize all
matrices accordingly while still keeping the small non-degeneracy of the KK masses, $M_{KK}\equiv M_{KK}^{\ell^{(1)}}\simeq M_{KK}^{\tau^{(1)}}\simeq0.87 
M_{KK}^{e^{(1)}}\simeq 0.95 M_{KK}^{\mu^{(1)}} \simeq 1.08 M_{KK}^{\tilde{e}^{(1)}}$. Finally, $v_\Phi^{4D}/\Lambda_{IR}\simeq0.27$ and  we assign $c_\ell\simeq c_\tau\simeq 0.51$, $c_e\simeq0.787$, $c_\mu\simeq0.63$, $y_e\simeq 1.2$, $y_\mu\simeq1.55$ and $y_\tau\simeq 1.7$ to reproduce the charged lepton masses with an IR scale $\Lambda_{IR}\simeq 1.5$\,TeV. The resulting masses are
$m_e=0.511$\,MeV, $m_\mu=105.7$\,MeV and $m_\tau=1.77$\,GeV.

 The
reason fermion KK mixing is so important for off diagonal $Z$
couplings is the presence of opposite chirality $(--)$ KK states and ``fake" custodial partners, which are
the $SU(2)_{R}$ partners of $\nu_R$ (Eq.~(\ref{FermionContent})) with $(-+)$ and
$(+-)$ boundary conditions. Recall that a 5D fermion corresponds
to two chiral fermions in 4D. As a result we will have LH states
which couple to the $Z$ boson as RH states (e.g. $\ell_R^{(1_{--})}$) and vice versa (e.g. $\tilde{e}_L^{(1_{+-})}$). As a result, the
$Z$ coupling matrices, $A_{L,R}(Z)$, which are proportional to the identity matrix in the
ZMA (up to small deviations induced by 
Eq.~(\ref{GaugeDeviation})), contain instead both types of (diagonal)
entries $(g_L^Z)_{SM},(g_R^Z)_{SM}$ and will thus acquire non vanishing off
diagonal elements, once rotated to the common mass basis of the KK and
zero mode fermions.

 Unfortunately, the three generation charged lepton (KK+zero) mass matrix can only be diagonalized numerically,
due to its dimension and the large number of input parameters. The  numerical diagonalization of this matrix is quite involved due to the presence of large hierarchies, ranging from $m_e$ to $M_{KK}$. It is thus essential to obtain an analytical approximation for  $(\delta g_{L,R}^{\mu e})_{KK}$, to get a control over the numerical results. For this purpose, we follow \cite{BurasKK} and use an effective field theory approach, where all heavy modes are integrated out. Within this approximation the zero mode mass matrix is modified in the following way,

\beq (M_{00})^{KK}=M_{00}+M_{0k}M_{k}^{-1}M_{kj}M_{j}^{-1}M_{j0}-\frac{1}{2}\left[M_{0k}M_k^{-2}M_{0k}^\dagger M_{00}+M_{00}M_{k0}^\dagger M_{k}^{-2}M_{k0}\right]+...\,,\label{MassBuras}\eeq

\noindent where the first correction originates from the pure heavy (KK) mass terms, while the second
correction containing $M_{00}$ is generated by the redefinitions of the light (SM) charged lepton fields, which are required to bring the kinetic terms into their canonical form \cite{BurasKK}. Notice that $M_{00}$ is precisely the ZMA mass matrix given in Eq.~(\ref{MDeltaM}), upon the identification $y_{e,\mu,\tau}^{4D}\equiv2 y_{e,\mu,\tau}F_\ell F_{e,\mu,\tau}(v_\Phi^{4D}/\Lambda_{IR})r_{00}^{e,\mu,\tau}$. In Eq.~(\ref{MassBuras}) the zero+KK mass matrix of Eq.(\ref{MFullD}) is decomposed into its ``generational" $3\times 3$ building blocks,  such that $M_{kj}$ is obtained by taking the $(kj)$ element of each of the $4\times 4$ block matrices in Eq.~(\ref{MFullD}). To simplify the notation, we label the zero and KK modes in the basis vectors of Eq.~(\ref{M4KK}) as $(0,1,2,3)$.  Observing Eq.~(\ref{M4KK}), we immediately realize that 
$M_{01}=M_{20}=M_{30}=M_{23}=M_{32}=0$, $M_{11,33}=\mathbbm{1}\cdot M_{KK}^{\ell^{(1)},\tilde{e}^{(1)}} $ and $M_{22}={\rm diag} (M_{KK}^{e^{(1)},\mu^{(1)},\tau^{(1)}})$.

The modified  mass matrix of Eq.(\ref{MassBuras})  induces  corrections to the left and right handed Z coupling matrices for the zero modes, $[A_{L,R}(Z)]_{00}$, where the first non vanishing corrections can be thought of as coming from placing two KK-zero mass insertions on the fermionic legs of a $Zf^{(0)}_{L,R}f^{(0)}_{L,R}$ vertex. The corrected coupling matrices are given by \cite{BurasKK}

\bea \left[A_L(Z)\right]_{00}^{KK}\!\!\!&=&\!\!\! \left[ A_L(Z)\right]_{00}+M_{0k}M_{k}^{-2}[A_L(Z)]_{kk}M_{0k}^\dagger\nonumber\\ \!\!\!&-&\!\!\!\frac{1}{2}\left[M_{0k}M_k^{-2}M_{0k}^\dagger\left[A_L(Z)\right]_{00} +\left[A_L(Z)\right]_{00}M_{00}M_{0k} M_{k}^{-2}M_{0k}^\dagger\right]+...\,,\label{gLdev}\\
 \left[A_R(Z)\right]_{00}^{KK}\!\!\!&=&\!\!\! \left[ A_R(Z)\right]_{00}+M_{k0}^\dagger M_{k}^{-2}[A_R(Z)]_{kk}M_{k0}\nonumber\\ \!\!\!&-&\!\!\!\frac{1}{2}\left[M_{k0}^\dagger M_k^{-2}M_{k0}\left[A_R(Z)\right]_{00} +\left[A_R(Z)\right]_{00}M_{00}M_{k0}^\dagger M_{k}^{-2}M_{k0}\right]+...\,.\label{gRdev} \eea

In the above equations, the corrected coupling matrices $\left[A_{L,R}(Z)\right]_{00}^{KK}$ are  given in the interaction basis
and should be rotated to the physical basis by the usual transformation $(V_{L,R}^{\ell})^\dagger[A_{L,R}]_{00}^{KK}V_{L,R}^\ell$. To account solely for KK mixing effects we ignore the small deviations implied by Eq.(\ref{GaugeDeviation}), such that $[A_{L,R}(Z)]_{00}=(g_{L,R}^Z)^{SM}\cdot\mathbbm{1}$, $[A_{L,R}]_{11}=(g_L^Z)_{SM}\cdot\mathbbm{1}$ and $[A_{L,R}(Z)]_{22,33}=(g_R^Z)_{SM}\cdot\mathbbm{1}$. To account for gauge boson mixing effects, we simply replace the first (zero order) terms in Eqs.~(\ref{gLdev}) and (\ref{gRdev}) by $[A_{L,R}(Z)]_{00}^G$ of Eqs.~(\ref{gLgauge}) and (\ref{GaugeCorrMass}).

We are now ready to compute the corrections to $\delta (g^{\mu e}_{L,R})_{SM}=0$  from fermion KK mixing. Starting from $A_L(Z)$and using $M_{01}=0$ and  $[A_{L,R}(Z)]_{22,33}=(g_R^Z)_{SM}\cdot\mathbbm{1}$ we have:

\beq A_L(Z)=[A_L(Z)]_{00}+((g^Z_R)_{SM}-(g_L^Z)_{SM}))\left[M_{02}M_{22}^{-2}M_{02}^\dagger+M_{03}M_{33}^{-2}M_{03}^\dagger\right],\label{ZleftDev}\eeq

where $(g^Z_R)_{SM}-(g_L^Z)_{SM}=1/2$. The block matrices $M_{02}$ and $M_{03}$ inherit their structure from the 5D Yukawa Lagrangian
and are given by:

\beq
M_{02}=\left(\begin{array}{ccc} (\breve{y}_e+\breve{x}_1^\ell f_\chi^e)F_\ell r_{01}^{e} x & (\breve{y}_\mu+\breve{x}_2^\ell f_\chi^\mu) F_\ell r_{01}^\mu x &(\breve{y}_\tau+\breve{x}_3^\ell f_\chi^\tau) F_\ell r_{01}^\tau x
\\ \breve{y}_eF_e r_{01}^{e} x & \omega\breve{y}_\mu F_\ell r_{01}^\mu x &\omega^2\breve{y}_\tau F_\ell r_{01}^\tau x\\
(\breve{y}_e+x_1^e f_\chi^e) F_\ell r_{01}^{e} x & (\omega^2\breve{y}_\mu+\breve{y}_2^\ell f_\chi^\mu)  F_\ell r_{01}^\mu x &
(\omega\breve{y}_\tau+\breve{y}_3^\ell f_\chi^\tau)  F_\ell r_{01}^\tau x \end{array}\right),\eeq

\beq
M_{03}=\left(\begin{array}{ccc} (\breve{y}_\nu+\breve{\epsilon}_1) F_\ell r_{101}^{\nu} x & \breve{\epsilon}_2 F_\ell r_{101}^{\nu}x &(\breve{\epsilon}_3+\breve{\epsilon}_\chi)F_\ell r_{101}^{\nu}x
\\ \breve{\epsilon}_3 F_\ell r_{101}^{\nu}x & (\breve{y}_\nu+\breve{\epsilon}_1) F_\ell r_{101}^{\nu} x&\breve{\epsilon}_2 F_\ell r_{101}^{\nu}x\\(\breve{\epsilon}_2+\breve{\epsilon}_\chi) F_\ell r_{101}^{\nu}x & \breve{\epsilon}_3 F_\ell r_{101}^{\nu}x &
(\breve{y}_\nu+\breve{\epsilon}_1) F_\ell r_{101}^{\nu} x \end{array}\right).\eeq

We expect the most dominant contribution to come from the term proportional to $M_{03}M_{33}^{-2}M_{33}^\dagger$ induced by the interactions of the LH zero modes $\ell_L^{(0)_{e,\mu,\tau}}$ with the ``custodians" $(\tilde{e}_R^{(1^{-+})},\tilde{
\mu}_R^{(1^{-+})},\tilde{\tau}_R^{(1^{-+})})$. Since $\breve{\epsilon}_{1,2,3}$ come from the $Z_3$ preserving  5D operator $k^{-7/2}\bar{\ell}_LH\Phi^2\nu_R$, we expect them to be ``rotated away" with $V_L^\ell$ at $\mathcal{O}(\epsilon_{1,2,3,\chi},f_\chi)$ and indeed using $M_{33}=M_{KK}^{\tilde{e}^{(1^{+-})}}\cdot\mathbbm{1}$ and Eq.~(\ref{C3:VLQ}), we  factorize out $(F_\ell r_{101}^\nu)^2 v^2/(M_{KK}^{\tilde{e}^{(1^{+-})}})^{2}$ and obtain:
\beq (V_L^\ell)^\dagger M_{03}M_{33}^{-2}M_{03}^\dagger V_L^\ell\propto\!\left(\!\!\begin{array}{ccc}|\breve{y}_\nu|^2+\frac{4}{3}{\rm Re}(\breve{y}_\nu\breve{\epsilon}_\chi) & -\frac{2\omega}{3}{\rm Re}(\breve{y}_\nu\breve{\epsilon}_\chi) &-\frac{2\omega^2}{3}{\rm Re}(\breve{y}_\nu\breve{\epsilon}_\chi) \\ -\frac{2\omega^2}{3}{\rm Re}(\breve{y}_\nu\breve{\epsilon}_\chi) & |\breve{y}_\nu|^2-\frac{2}{3}{\rm Re}(\breve{y}_\nu\breve{\epsilon}_\chi) &  \frac{4\omega}{3}{\rm Re}(\breve{y}_\nu\breve{\epsilon}_\chi) \\  -\frac{2\omega}{3}{\rm Re}(\breve{y}_\nu\breve{\epsilon}_\chi) & \frac{4\omega^2}{3}{\rm Re}(\breve{y}_\nu\breve{\epsilon}_\chi) & |\breve{y}_\nu|^2-\frac{2}{3}{\rm Re}(\breve{y}_\nu\breve{\epsilon}_\chi) \end{array}\!\!\right).\label{M03correction}\eeq

\noindent The above expression is already given in the mass basis and to get the contribution to the $Z\mu e$ coupling we simply take the $(12)$ element of the matrix in Eq.~(\ref{M03correction}), recalling the prefactors:
\beq
\delta (g_L^{\mu e})_{KK}^{\tilde{e}^{(1)}}\simeq\left|\frac{-\omega}{3}{\rm Re}(\breve{y}_\nu\breve{\epsilon_\chi})(F_\ell r_{101}^\nu)^2\frac{v^2}{(M_{KK}^{\tilde{e}^{(1^{+-})}})^2}\right|\approx9\times10^{-8}, \label{gLFinalCust}\eeq

\noindent where we used $r_{101}^\nu\simeq0.72$, $F_\ell\simeq0.1$ and set $|y_{\nu,\epsilon_\chi}|=1$. We now proceed to obtain the correction coming from the $M_{02}M_{22}^{-2}M_{02}^\dagger$ term in Eq.~(\ref{ZleftDev}). We expect this term to generate a suppressed contribution due to the approximate alignment of $M_{02}$ and $M_{00}$. Notice that due to the degeneracy of $c_{\ell}$ , we can write $M_{02}$ as 
\beq M_{02}=M_{00} \,{\rm diag}\left[F_{e,\mu,\tau}^{-1}(r_{00}^{e,\mu,\tau})^{-1}r_{01}^{e,\mu,\tau}\right],\label{M02simple} \eeq
using this compact form it is straightforward to transform this correction to the mass basis:
\bea (V_L^\ell)^\dagger M_{02}M_{22}^{-2}M_{02}^\dagger V_L^\ell \!\!&=&\!\! (V_L^\ell)^\dagger M_{00}V_R^\ell (V_R^\ell)^\dagger \,{\rm diag}\!\left(
\frac{(r_{01}^{e,\mu,\tau})^2}{\left(F_{e,\mu,\tau}r_{00}^{e,\mu,\tau}M_{KK}^{e^{(1)},\mu^{(1)},\tau^{(1)}}\right)^{2}}\right)\! V_R^\ell(V_R^\ell)^\dagger M_{00}^\dagger V_L^\ell\nonumber\\ \!\!&=&\!\!{\rm diag}(m_{e,\mu,\tau})(V_R^\ell)^\dagger\,{\rm diag}\!\left(
\frac{(r_{01}^{e,\mu,\tau})^2}{\left(F_{e,\mu,\tau}r_{00}^{e,\mu,\tau}M_{KK}^{e^{(1)},\mu^{(1)},\tau^{(1)}}\right)^{2}}\!\right)V_R^\ell\,{\rm diag}(m_{e,\mu,\tau})\nonumber\\ \!\!&=&\!\! m_{e_i}m_{e_j}\sum_{n=1}^3(V_R)_{ni}^*(V_R)_{nj}\!\left(
\frac{(r_{01}^{e,\mu,\tau})^2}{\left(F_{e,\mu,\tau}r_{00}^{e,\mu,\tau}M_{KK}^{e^{(1)},\mu^{(1)},\tau^{(1)}}\right)^{2}}\right)_{nn}\!\label{ZM02Dev}.\eea
It is now straight forward to estimate  $(\delta g_L^{\mu e})_{KK}^{e_i^{(1)}}$ by taking the  $(12)$ element from the above expression. The dominant term acquires the form:
\beq (\delta g_L^{\mu e})_{KK}^{e_i^{(1)}}\simeq\frac{m_em_\mu\Delta_1^\ell(r_{01}^\mu)^2}{\left(F_\mu r_{00}^\mu M_{KK}^{\mu^{(1)}}\right)^2} \approx 6\times 10^{-11},\label{gLFinal}\eeq

\noindent where we set all Yukawas to 1 in magnitude and use $F_\mu\simeq 0.004$, $r_{01}^\mu\simeq 0.8$ and $r_{00}^\mu\simeq 0.87$.
This concludes the calculation of KK fermion mixing contribution to $\delta g_L^{\mu e}$.

 We now turn to the RH couplings, in which case we have only one term contributing, proportional to $M_{10}^\dagger M_{11}^{-2}M_{10}$ and thus Eq.~(\ref{gRdev} simplify to:
\beq A_R(Z)=[A_R(Z)]_{00}+((g_L^Z)_{SM}-(g_R^Z)_{SM})\left[M_{10}^\dagger M_{11}^{-2}M_{10}\right],\label{gRdevFinal}\eeq

\noindent where $(g_L^Z)_{SM}-(g_R^Z)_{SM}=-1/2$, $M_{11}=M_{KK}^{\ell^{(1)}}\cdot\mathbbm{1}$ and the various entries of $M_{10}$ are given by

\beq
M_{10}=\left(\begin{array}{ccc} (\breve{y}_e+\breve{x}_1^\ell f_\chi^e)F_e r_{10}^{e} x & (\breve{y}_\mu+\breve{x}_2^\ell f_\chi^\mu) F_\mu r_{10}^\mu x &(\breve{y}_\tau+\breve{x}_3^\ell f_\chi^\tau) F_\tau r_{10}^\tau x
\\ \breve{y}_eF_e r_{10}^{e} x & \omega\breve{y}_\mu F_\mu r_{10}^\mu x &\omega^2\breve{y}_\tau F_\tau r_{10}^\tau x\\
(\breve{y}_e+x_1^e f_\chi^e) F_e r_{10}^{e} x & (\omega^2\breve{y}_\mu+\breve{y}_2^\ell f_\chi^\mu)  F_\mu r_{01}^\mu x &
(\omega\breve{y}_\tau+\breve{y}_3^\ell f_\chi^\tau)  F_\tau r_{10}^\tau x \end{array}\right).\eeq

\noindent Once again, the degeneracy of $c_\ell$  implies that $M_{10}$ is approximately aligned with $M_{00}$ and can be written in the following way:
\beq M_{10}=M_{00}\,{\rm diag}(r_{10}^{e,\mu,\tau}/r_{00}^{e,\mu,\tau})F_\ell^{-1}. \label{M01simple}\eeq

\noindent The above equation simplifies the transformation of Eq.~(\ref{gRdevFinal}) to the mass basis and we get:
\bea (V_R^\ell)^\dagger M_{10}^\dagger M_{11}^{-2}M_{10}V_R^\ell\!\!&=&\!\!(V_R^\ell)^\dagger \,{\rm diag}\left(\frac{r_{10}^{e,\mu,\tau}}{r_{00}^{e,\mu,\tau}}\right)\!V_R^\ell (V_R^\ell)^\dagger M_{00}^\dagger V_L^\ell\nonumber\\ \!\!&\cdot&\!\!(V_L^\ell)^\dagger M_{00} V_R^\ell(V_R^{\ell})^\dagger {\rm diag}\left(\frac{r_{10}^{e,\mu,\tau}}{\left(M_{KK}^{\ell^{(1)}}\right)^2r_{00}^{e,\mu,\tau}}\right)\! V_R^\ell  \label{gRdevFuck}. \eea

We use the same steps taken for the LH couplings in Eq.~(\ref{ZM02Dev}) to simplify the long expression in Eq.~(\ref{gRdevFuck}). The resulting expression for $[\Delta A_R]^{mass}\equiv (V_R^\ell)^\dagger M_{10}^\dagger M_{11}^{-2}M_{10}V_R^\ell$
reads,
\beq [\Delta A_R]^{mass}_{ij}=F_\ell^{-2}\left[\,\sum_{n,k,l=1}^{3}(V_R^\ell)_{ni}^*(V_R^\ell)_{nk}\,(m_{e_k})^2\,{\rm diag}\left(\frac{r_{10}^{e,\mu,\tau}}{r_{00}^{e,\mu,\tau}}\right)_{nn}(V_R^\ell)_{lk}^*(V_R^\ell)_{lj}\,{\rm diag}\left(\frac{r_{10}^{e,\mu,\tau}}{r_{00}^{e,\mu,\tau}}\right)_{ll}\,\right]. \label{gRFinal}\eeq

We realize that due to the degeneracy of $c_\ell$ the LH diagonalization matrices are absent in the above expression, implying that the off diagonal elements of $[\Delta A_R]^{mass}_{ij}$ will come from terms proportional to the NLO corrections of $V_R^\ell$ and will also consist of additional cancellation patterns, induced by the near degeneracy of the Higgs-flavon overlap correction factors $r_{10,00}^{e,\mu,\tau}$. Specifying to the RH $Z\mu e$ coupling, we extract the dominant terms and obtain

\bea (\delta g_R^{\mu e})_{KK}^{\ell^{(1)}}\simeq F_\ell^{-2}\left[\frac{m_\mu^2 \Delta_1^\ell r_{10}^{\mu}}{\left(M_{KK}^{\ell^{(1)}}\right)^2r_{00}^{\mu}}
\left(\frac{r_{10}^e}{r_{00}^e}-\frac{r_{10}^
\mu}{r_{00}^\mu}\right)\right]+...\approx  10^{-12}\,, \eea

\noindent where we have omitted terms, which are  at least $\mathcal{O}(f_\chi\approx 0.05)$ suppressed compared to those appearing in the above equation. To get the numerical estimation, we assigned all Yukawas to 1 and used $r_{00}^e\simeq0.895$, $r_{00}^\mu\simeq0.87$
$r_{00}^\tau \simeq 0.86$, $r_{10}^e\simeq0.754$, $r_{10}^\mu\simeq 0.74$ and $r_{10}^\tau\simeq0.73$. To summarize our analytical estimations we have:
\beq (\delta g_L^{\mu e})^{gauge}=0, \quad (\delta g_L^{\mu e})^{KK}\approx 9\times10^{-8}, \quad (\delta g_R^{\mu e})^{gauge}\approx 5\times 10^{-11}, \quad (\delta g_R^{\mu e})^{KK}\approx 10^{-12}.\label{FinalEstimationsZ}\eeq

\subsection{FCNC protection in the brane localized RS-A$_4$ setup}\label{Sec:Protection}
Since both the gauge and fermionic contributions, calculated in the previous sections, come from ``cross-brane" effects it seems
natural to believe that the brane localized RS-A$_4$ setup is completely protected from all sources of tree level FCNC. This point was discussed in \cite{Csaki:2008qq}, yet the fermion KK mixing effects, were not taken into account there. Indeed, in the brane localized case we have $V_R^\ell=\mathbbm{1}$, which together with the degeneracy of $c_\ell$ implies there can be no gauge boson mixing  induced contributions to tree level FCNC (Eq.~(\ref{GaugeCorrMass})). Similarly, the analytical approximations for $(\delta g_{L,R}^{\mu e})^{KK}$ (Eqs.~(\ref{gLFinalCust}), (\ref{gLFinal}) and (\ref{gRFinal})) suggest that these sources of FCNC are turned off as well. Going to higher order in the expansion of Eqs.~(\ref{gLdev}) and (\ref{gRdev}) won't do the trick since all blocks of the KK+zero mass matrix inherit their structure from the 5D Yukawa lagrangian, which is ``form diagonal" in the brane localized case. Due to the $Z_3$ preserving VEV of $\Phi$ (Eq.~(\ref{VEValignment}), the interactions with the custodians $(\tilde{e}^{(1)},\tilde{\mu}^{(1)},\tilde{\tau}^{(1)})$, which mimic the Dirac interactions in the neutrino sector, are also "form diagonalizable",   namely, diagonalized by  $U(\omega)$ (Eq.~(\ref{MDeltaM}))  and thus can not contribute to FCNC.

The absence of tree level FCNC is not surprising since the LH zero modes are degenerate and the RH zero modes require no rotation to be transformed into the mass basis. This means that one can always find a basis in which the kinetic and mass terms are simultaneously diagonal, which simply forbids FCNC. Nevertheless, we have shown in the context of one loop contributions to dipole operators in the quark sector \cite{A4CPV}, that the degeneracy between the KK masses, which is even stronger in the charged lepton sectors, implies that a non perturbative rotation of the degenerate KK blocks is needed before we can act with standard techniques of
non-degenerate perturbation theory. Such a rotation, mixes the KK states maximally with each other (generation+BC) and might still induce small contributions to tree level FCNC. To verify the extent to which such a situation contributes to FCNC we shall attempt
to analytically diagonalize the zero+first KK modes mass matrix. This seemingly hard task, shall be greatly simplified due to the magic properties of $U(\omega)$. We start by writing the $9\times9$ KK-KK block of the full mass matrix in terms of $3\times3$ building blocks,
\beq
{\bf M}_{KK}^{9\times9}\propto\left(\begin{array}{ccc} M_{KK}^{\ell^{(1)}}\cdot\mathbbm{1} & \sqrt{3}U(\omega){\rm diag}
(y_{e,\mu,\tau}r_{11}^{e,\mu,\tau})v & \hat{Y}_{KK}^\nu r_{111}^\nu \,v\\ \sqrt{3}U(\omega){\rm diag}
(y_{e,\mu,\tau}^*r_{22}^{e,\mu,\tau})v & {\rm diag}\left(M_{KK}^{e^{(1)},\mu^{(1)},\tau^{(1)}}\right) & 0\\ 
(\hat{Y}_{KK}^\nu)^\dagger r_{222}^\nu \,v & 0 & M_{KK}^{\tilde{e}^{(1^{+-})}}\cdot\mathbbm{1} \end{array}\right),\label{KK9X9}\eeq

\noindent where $v=174$\,GeV is the Higgs VEV and  $\hat{Y}_\nu^{KK}$ is a dimensionless coupling matrix, inheriting its structure from the Dirac neutrino mass matrix (Eq.~(\ref{DiracMassMatrix})) and is given by:
\beq \hat{Y}_{\nu}^{KK}=\left(\begin{array}{ccc} y_\nu+\epsilon_1 & \epsilon_2 &\epsilon_3\\\epsilon_3 &y_\nu+\epsilon_1 &\epsilon_2\\
\epsilon_2 &\epsilon_3 &y_\nu+\epsilon_1\end{array}\right).\label{NuKKCouplingMatrix}\eeq

Because of the symmetry between the $\epsilon_{2,3}$ entries, which is inherited from the $Z_3$ preserving VEV of $\Phi$, we expect $\hat{Y}_\nu^{KK}$ to be diagonalized by $U(\omega)$ and indeed we realize that:

\beq \hat{Y}_\nu^{KK}= U(\omega)\underbrace{\left(\!\!\begin{array}{ccc} y_\nu+\epsilon_1+\epsilon_2+\epsilon_3 &0 &0\\0&y_\nu+\epsilon_1+\omega\epsilon_2+\omega^2\epsilon_3&0\\0&0&y_\nu+\epsilon_1+\omega^2\epsilon_2+\omega\epsilon_3\end{array}\!\!\right)}_{\hat{Y}_\nu^{KK(diag.)}}\left(U(\omega)\right)^\dagger\label{NuKKCouplingMatrixDiag}.\eeq

Observing the above features of $U(\omega)$, we realize that the first trivial rotation we can impose on the KK mass matrix
of Eq.~(\ref{KK9X9}), will be block diagonal, where each $3\times3$ building block is identified with either $U(\omega$ or the identity matrix. Such a rotation will leave the $Z$ coupling matrices, $A_{L,R}(Z)$ unchanged, since each $3\times3$ block in $A_{L,R}(Z)$ is proportional to the identity. In particular consider, the matrix $U_{L[R]}^{12\times12}\!={\rm diag}\,[U(\omega)[\mathbbm{1}],\mathbbm{1},U(\omega),U(\omega)]$. Acting with $U_L^{12\times12}$ on ${\bf \hat{M}}^{\ell}_{Full}$ will make each $3\times3$ block generation diagonal, which means that the nine dimensional degenerate subspace, which had to be diagonalized has splitted into three, according to the ``type"\,(BC) of KK modes. Assuming the degeneracy of KK masses,  the rotated KK+zero mass matrix, ${\bf \hat{M}}_{Full}^{\ell(U)}$ is proportional to:
\beq \left(\begin{array}{cccc} {\rm diag}(\breve{y}_{e,\mu,\tau}F_\ell F_{e,\mu,\tau}r_{00}^{e,\mu,\tau}) & 0 & {\rm diag}(\breve{y}_{e,\mu,\tau})F_{e,\mu,\tau}(r_{01}^{e,\mu,\tau})) & \breve{Y}_{KK}^{\nu(diag.)} F_\ell r_{101}^\nu
\\{\rm diag}(\breve{y}_{e,\mu,\tau} F_{e,\mu,\tau}r_{10}^{e,\mu,\tau})&1/x&{\rm diag}(\breve{y}_{e,\mu,\tau})r_{11}^{e,\mu,\tau}&\breve{Y}_{KK}^{\nu(diag.)} F_\ell r_{111}^\nu\\0&{\rm diag}(\breve{y}_{e,\mu,\tau}^*)r_{22}^{e,\mu,\tau}&1/x&0\\0&\breve{Y}_{KK}^{\nu(diag.)\dagger} F_\ell r_{222}^\nu&0&1/x\end{array}\right),\label{MFULLUU}\eeq

\noindent where $x=v/M_{KK}$. We realize that all $3\times3$ blocks of the above matrix are diagonal, namely  we can treat each blocks as a set of three complex numbers, without worrying about the inversion of parametric $3\times3$ matrices. Recalling  the independence of $U(\omega)$ from the LO Yukawa couplings, we realize that the $9\times9$ KK block of the matrix in Eq.~(\ref{MFULLUU}) has the characteristic form:
\beq \left({\bf \hat{M}}_{Full}^\ell(U)\right)_{9\times 9}^{KK}\propto M_{KK}^{ch.}\equiv\left(\begin{array}{ccc} 1/x&A&B\\aA^*&1/x&0\\bB^*&0&1/x
\end{array}\right),\label{MKKblocktype}\eeq

\noindent where $A\equiv{\rm diag}(\breve{y}_{e,\mu,\tau})r_{11}^{e,\mu,\tau}$, $B=\breve{Y}_{KK}^{\nu(diag.)} F_\ell r_{111}^\nu$ and $\hat{a}(\hat{b})={\rm diag}(r_{22(222)}^{e,\mu,\tau(\nu)}/r_{11(111)}^{e,\mu,\tau(\nu)})$.
Using $r_{11}^e\simeq 0.763$, $r_{11}^\mu\simeq 0.758$, $r_{11}^\tau\simeq 0.753$, $r_{111}^\nu\simeq0.749$, $r_{22}^e\simeq 0.383$ and $r_{22}^\mu\simeq 0.367$, $r_{22}^\tau\simeq 0.357$ and $r_{222}^\nu\simeq 0.19$, we realize that $\hat{a}\simeq \hat{b}\simeq1/2\cdot\mathbbm{1}$. The resulting $M_{KK}^{ch.}$ can be diagonalized analytically by the  transformation, $M_{KK}^{ch.(diag.)}=(V_L^{KK})^\dagger M_{KK}^{ch.}V_R^{KK}$, where  $V_{L,R}^{KK}$ are given by:
\beq V_{L,R}^{KK}=\left(\begin{array}{ccc}0&-\frac{e^{\pm i\theta_B}}{\sqrt{2}}&\frac{e^{\pm i\theta_B}}{\sqrt{2}}\\
-\frac{|B|e^{\pm i(\theta_B-\theta_A)}}{\sqrt{|A|^2+|B|^2}}&\frac{|A|e^{\pm i(\theta_{B}-\theta_A)}}{\sqrt{2(|A|^2+|B|^2)}}& \frac{|A|e^{\pm i(\theta_B-\theta_A)}}{\sqrt{2(|A|^2+|B|^2)}}\\\frac{|A|}{\sqrt{|A|^2+|B|^2}}&\frac{|B|}{\sqrt{2(|A|^2+|B|^2)}}&\frac{|B|}{\sqrt{2(|A|^2+|B|^2)}}\end{array}\right),\eeq

\noindent and where $M_{KK}^{ch.(diag.)}={\rm diag}\left(1, 1-3/2\sqrt{|A|^2+|B|^2}, 1+3/2\sqrt{|A|^2+|B|^2}\right)$. We are now left only with the non vanishing zero-KK blocks\,--\, they are mixtures of the flavor diagonal blocks, ${\rm diag}(\breve{y}_{e,\mu,\tau}F_{e,\mu,\tau(\ell)}r_{01(10)}^{e,\mu,\tau})$ and $\hat{Y}_{KK}^{\nu(diag.)}F_\ell r_{101}^\nu$. Therefore, the remaining (standard) perturbative diagonalization of $M_{KK}^{ch.(diag.)}$ won't contribute
to FCNC, to all orders. If we will consider the explicit values of the overlap correction factors, $r_{nm}$ above, flavor violating couplings may be generally generated at $\mathcal{O}(v^4/M_{KK}^4)$, but will be further suppressed by quartic differences of the nearly degenerate $r_{nm}$. Such contributions are negligibly small $(\mathcal{O}(10^{-18}))$ and will thus not be further discussed. Finally, it is important to stress that the above diagonalization scheme can be easily generalized to include an arbitrary number of KK modes and will yield the same results in this case.

\subsection{The RS-A$_4$ contributions to $\mu\to 3e$ and $\mu\to e$ conversion}\label{Sec:Znumerical}
The effective Lagrangian for $\mu\to 3e$ deacy and $\mu\to e$ conversion in  nuclei is written in terms of four fermion operators. These lepton flavor-violating Fermi operators are traditionally parametrized as \cite{LFVchina}
\begin{eqnarray}
{\cal L}&=& \frac{4 G_F}{\sqrt{2}} \left[ g_3 (\bar{e}_R \gamma^\mu \mu_R)(\bar{e}_R \gamma_\mu e_R) +g_4 (\bar{e}_L \gamma^\mu \mu_L)(\bar{e}_L \gamma^\mu e_L)+g_5 (\bar{e}_R \gamma^\mu \mu_R)(\bar{e}_L\gamma_\mu e_L) \right.
\nonumber \\ && \left.+g_6 (\bar{e}_L \gamma^\mu \mu_L)(\bar{e}_R \gamma_\mu e_R) \right]
 +\frac{G_F}{\sqrt{2}} \bar{e} \gamma^\mu (v-a \gamma_5) \mu \sum_q \bar{q} \gamma_\mu (v^q-a^q \gamma_5) q,
 \label{eq:effLag}
\end{eqnarray}
where in the normalization we use, $v^q=T_3^q-2 Q^q \sin^2\theta$. The axial coupling to quarks, $a^q$, vanishes in the dominant contribution coming from coherent scattering off the nucleus. The $g_{3,4,5,6}$ are responsible for $\mu\to 3e$ decay, while the $v,a$ are responsible for $\mu\to e$ conversion in nuclei. The rates are given by (with the conversion rate normalized to the muon capture rate):
\begin{align}
\text{Br}(\mu\to 3e)&= 2 (g_3^2+ g_4^2)+g_5^2+g_6^2\ ,  \label{eq:Br:mu3e}\\
\text{Br}(\mu\to e)&= \frac{ p_e E_e G_F^2 F_p^2 m_\mu^3 \alpha^3 Z_{eff}^4}{\pi^2 Z \Gamma_{\text{capt}}} Q_N^2  (v^2+a^2),\label{eq:Br:mue}
\end{align}
where the parameters for the conversion depend on the nucleus and are calculated in the Feinberg-Weinberg approximation 
\cite{Feinberg} and we write the charge for a nucleus with atomic number $Z$ and neutron number $N$ as
\begin{align}
    Q_N = v^u (2Z+N)+v^d(2N+Z).
\end{align}
\noindent The most sensitive experimental constraint, ${\rm Br}(\mu\to e)\lesssim 6\times 10^{-13}$, comes from muon conversion in ${}_{22}^{48}\text{Ti}$ \cite{sindrumII}, for which
\begin{equation}
    E_e\sim p_e \sim m_\mu,
    \quad\quad\quad\quad
    F_p \sim 0.55,
    \quad\quad\quad\quad
    Z_{\text{eff}}\sim 17.61,
    \quad\quad\quad\quad
    \Gamma_{\text{capt}} \sim 2.6 \cdot \frac{10^6}{\text{s}}.
\end{equation}

Using these couplings one can estimate the coefficients of the 4-Fermi operators in (\ref{eq:effLag}),
\begin{equation}
    g_{3,4} = 2(\delta g_{L,R}^{\mu e})^2\quad\quad\quad\quad\quad\quad
    g_{5,6} = 2\delta g_{L}^{\mu e}\delta g_{R}^{\mu e}\quad\quad\quad\quad\quad\quad
    (v\pm a) = 2 \delta g_{L,R}^{\mu e},
    \label{eq:tree:Z:couplings}
\end{equation}
where $\delta g_{L,R}^{\mu e}$ contains both the gauge and fermionic contributions to the  $Z\mu e$ coupling.

\begin{center}
\begin{figure}
\includegraphics[width=7.45cm,scale=0.9]{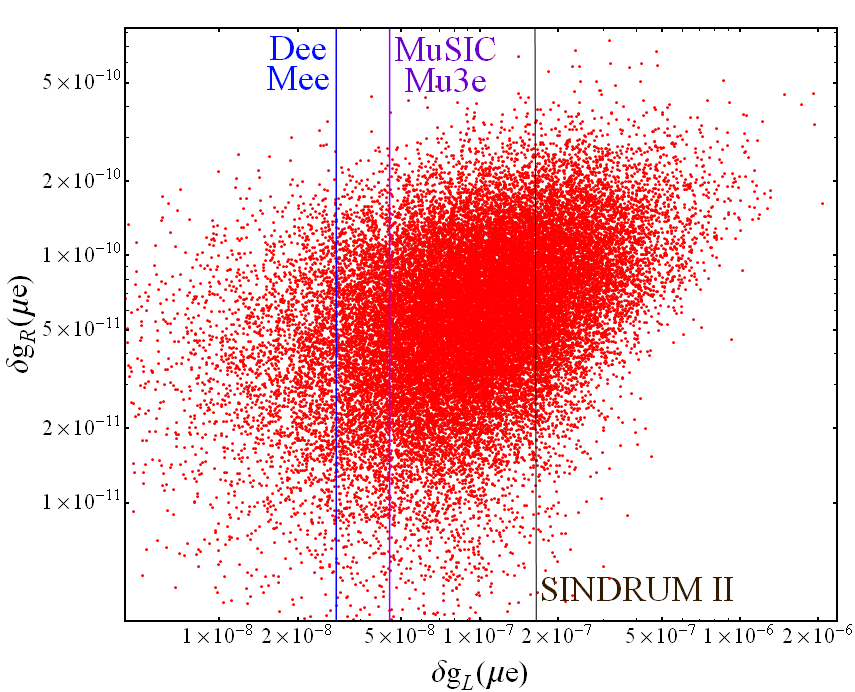}\qquad\qquad\!\!\!\!\!\!\!\!
\includegraphics[width=7.55cm]{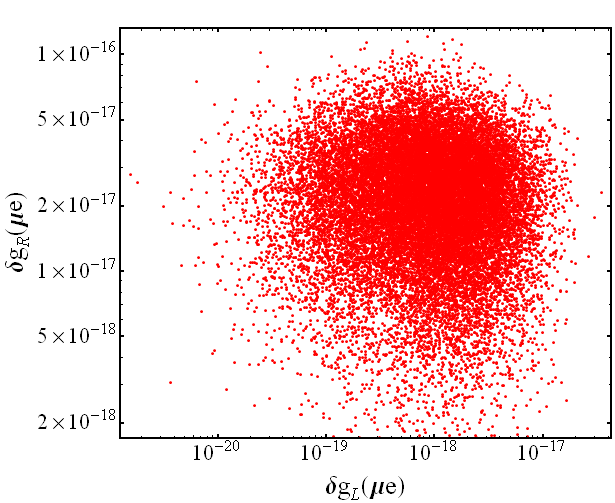}\caption{\it
The RS-A$_4$ predictions for the anomalous LH and RH $Z\mu e$
couplings in the presence (left) or absence (right) of cross brane
interactions. Each point represents the contributions coming from
both gauge boson mixing and KK fermion mixing. including all
dominant higher order and cross talk effects. The dashed lines
represent the maximum sensitivities of past, present and future
LFV experiments taking place at FERMILAB and
J-PARC.}\label{Fig:LFVscatter}
\end{figure}
\end{center}

Recalling our collective estimations of the gauge and fermionic contributions (Eq.~(\ref{FinalEstimationsZ})), obtained for Yukawa couplings with magnitude 1 and random phase, we have $(\delta g_{L(R)}^{\mu e})_{G+KK}\approx8\times10^{-8}\,(5\times10^{-11})$.
We expect the results of a numerical scan over the magnitudes and phases of the  Yukawa couplings to be centered around these values
if the effective field theory approach we used (Eq.~(\ref{MassBuras})) is reliable. To find the exact result we perform a numerical diagonalization of ${\bf \hat{M}}_{Full}^\ell$ and use the resulting diagonalization matrices, 
${\bf V_{L,R}^{KK(full)}}$ to rotate the $12\times12$ coupling matrices $A_{L,R}(Z)={\rm diag}\left([A_{L,R}(Z)]_{00,11,22,33}\right)$ to the zero+KK mass basis, where we have $(\delta g_{L,R}^{Z\mu e})_{mass}=[A_{L,R}(Z)]_{\mu^{(0)} e^{(0)}}^{mass}$. Once we obtain $(\delta g_{L,R}^{Z\mu e})_{mass}^{RS-A_4}$ we perform a matching to the operators
contributing to $\mu\to e,3e$ (Eq.~(\ref{eq:effLag})) using Eqs.~(\ref{eq:Br:mu3e}) and (\ref{eq:Br:mue}). Using the points satisfying the $3\sigma$ bounds on the neutrino mixing angles from the scan of Sec.~\ref{Sec:NUanalyze}, we generate a  sample of 40000 points, consisting of the $\epsilon_{2,3,4,\chi}$, $\tilde{x}_{2,3}^{\ell}$ and $\tilde{y}_{2,3}^\ell$ parameters. The distribution of the values of each of these parameters, within the sample, is practically indistinguishable from the original distribution from which it was taken from. Namely, all parameters are still complex numbers with random phases and magnitudes normally distributed around 1 with standard deviation 0.5. Since $BR(\mu\to e,3e)\propto
(g_{L,R}^{Z\mu e})^2$ (Eqs.~(\ref{eq:Br:mu3e})\,--\,(\ref{eq:Br:mue})) and $\delta g_L$ generally dominates, we can
plot the constraints coming from the various cLFV experiments as
vertical lines in the $\delta g_L^{\mu eZ}-\delta g_R^{\mu eZ}$
plane. The results are depicted in Fig.~\ref{Fig:LFVscatter} (left). The results of a separate scan, independent of constraints from the neutrino sector, reveal negligible differences and are thus not discussed separately. By observing Eqs.~(\ref{T12})\,--\,(\ref{T23}) we recall that all mixing angles receive contributions from four NLO Yukawa couplings in the charged lepton sector $\tilde{x}_{2,3}^\ell$ and $\tilde{y}_{2,3}^\ell$. Consequently, there are simply too many input parameters of $\mathcal{O}(1)$ and random phases, which can't be significantly constrained by the experimental data. By observing Fig.~\ref{Fig:LFVscatter} we realize that the numerical results are in an excellent agreement with the estimations of Eq.~(\ref{FinalEstimationsZ}). This strengthens our confidence in the effective field theory approach adopted from \cite{BurasKK}. The current
bound from SINDRUM II \cite{sindrumII} ``eliminates" around 35\% of the
cross-brane RS-A$_4$ setup and is easily and naturally satisfied for $\Lambda_{IR}\simeq1.5$\,TeV. The correlation between the LH and RH $Z\mu e$ couplings is attributed to the fact that they are both generated by cross brane effects, induced by $v_\chi$ (Eq.~(\ref{VEVprofile})).
%where the left panel corresponds to an analysis of all possible
%contributions and in the right panel cross-brane effects are neglected.
 When cross brane effects are neglected, the $Z\mu e$ couplings are negligible (right plot of Fig.~\ref{Fig:LFVscatter}), while corrections to the TBM pattern can only come from higher order corrections to the
heavy Majorana and Dirac mass matrices. Recall that in this case, achieving
$\theta_{13}\sim\theta_C/\sqrt{2}$ is slightly less natural from the 5D theory point of view (requires $|y_{\epsilon_4}|\sim 6$) (Sec.~\ref{Sec:Nubrane}). 
We conclude that if no $\mu\to e,3e$ events will be actually
observed in Mu3e, MuSIC and Dee-Mee the ``cross talk" RS-$A_4$
model will be severely constrained and less appealing. On the other hand, the predictions for the brane localized realizations of
RS-A$_4$ are  extremely far  from the reach of Mu2e\,\cite{Mu2e}, COMET\,\cite{COMET}, PRIME\,\cite{PRIME} and other future
experiments. Most probably, the same situation will not hold true for the $\mu\to e\gamma$ decay induced at the one-loop level. The current upper bound is ${\rm Br}(\mu\to e\gamma)\lesssim 6\times 10^{-13}$\,\cite{MEG} and is expected to improve by roughly two orders of magnitude in the next few years. This concludes our analysis of cLFV in RS-A$_4$ setups.

%\newpage

%\begin{fmffile}{fgraphs}
%\fmfcmd{% 
 %vardef cross_bar (expr, p, len, ang) =
%((-len/2,0)--(len/2,0))
%rotated (ang + angle direction length(p/2) of p)
%shifted point length(p/2) of p
%enddef;
%style_def crossed expr p =
%cdraw p;
%ccutdraw cross_bar (p, 5mm, 45);
%ccutdraw cross_bar (p, 5mm, -45)
%enddef;}

%\begin{center}
%\begin{fmfchar*}(40,25)
% \fmfleft{em,ep}
% \fmf{fermion}{em,Zee,ep}
% \fmf{crossed,label=$Z$}{Zee,Zff}
%\fmfcmd{tfermion}--{ep,f}
 %\fmf{fermion}{fb,Zff,f}
 %\fmfright{fb,f}
 %\fmfdot{Zee,Zff}
 %\end{fmfchar*}\end{center}\end{fmffile}

\section{Conclusions}\label{Sec:Conclusions}
\noindent In this work we have studied  the charged lepton and neutrino sectors of RS-A$_4$ (seesaw I) models
excluding/including cross brane effects. For the neutrino sector, we have shown that  cross-talk
operators in the charged lepton sector and brane localized higher order corrections to the (heavy) Majorana 
and Dirac mass matrices induce significant deviations from  TBM neutrino mixing, such that the experimental bounds can be rather easily satisfied. The mild differences between the normal and inverted hierarchy cases come from the terms proportional 
to $\epsilon_{2,3,4}$ in Eqs.~(\ref{T12})\,--\,(\ref{T23}), associated with brane localized higher dimensional operators in the neutrino sector . The success rates for satisfying the $3\sigma$ bounds from all mixing angles simultaneously were shown to be 
$\xi_{NH(IH)}\simeq10\%(12.5\%)$, which is slightly higher than those associated with ``typical" A$_4$ seesaw models \cite{FeruglioNew} and with anarchic models \cite{FeruglioAnarchy}. Therefore, despite the fact that the recent measurements of $\theta_{13}\approx\lambda_C/\sqrt{2}$ and  the growing indications for the non maximality of $\theta_{23}$ \cite{Fogli,Tortola}  deviate significantly from the TBM  pattern, models based on TBM mixing at LO with NLO corrections coming from both the neutrino and charged lepton sectors are still  viable in explaining the neutrino mixing angles. The advantage of the RS-A$_4$ model remain in its relative simplicity and the fact that NP contributions to cLFV processes are very suppressed. When specializing to the brane localized case, the resulting simplifications fix $\theta_{13}$ by a single (real) parameter and the remaining two  parameters are first assumed to be equal up to a phase to look for  possible correlations among the predictions. Indeed, we find a very strong correlation  between the value of $\theta_{13}$ (close to $9^\circ$) and the  the deviation from maximality of $\theta_{23}$ towards the first (second) octant for the ``positive"(``negative") cases, as can be seen in Fig.~\ref{NUAngles32}. In the same context it is worth mentioning a slightly different approach \cite{MuChun2012,Ma2012}, in which the parameter space of the model is narrowed down to include only the VEVs of four flavons $(\3,\s,\spr,\sppr)$, corresponding to four input parameters in the $Z_{2,3}$ preserving cases. These parameters are thus over-constrained by the existing neutrino measurements and the existence of a non trivial solution to this system of constraints reflects the suitability of A$_4$ as a flavor symmetry for the lepton sector. Another interesting example is the so called ``special" A$_4$ models described in \cite{FeruglioNew}, where one is able to obtain a significantly better  success rate $\xi_{A_4}^{spc.}\simeq 50\%$. An example for such a model can be found in \cite{LinA4}. 

\noindent Turning back to the charged lepton sector, we have studied in detail the way in which KK mixing effects of gauge bosons and fermions generate flavor violating $Z$ couplings in charge of tree level FCNC. Due to the form diagonalizability of the LO mass matrices, the anomalous $Z$ couplings are generated at the NLO by ``cross-brane" effects which break A$_4$ completely. We have obtained analytical approximations to the anomalous $Z\mu e$ coupling using an effective field theory approach, in which all KK modes are integrated out. It was shown that the structure of the (non-custodian) corrections to the LH and RH $Z$ coupling matrices (Eqs. (\ref{ZM02Dev}) and (\ref{gRFinal})) are governed by the RH diagonalization matrices,  implying a further suppression of $f_\chi^\mu(m_e/m_\mu)$ compared to the anarchic case and thus a partial protection against tree level FCNC. The results of the exact numerical diagonalization are in excellent agreement with the analytical approximations and  are depicted in the left plot Fig.~\ref{Fig:LFVscatter}. We realize that the current upper bounds from SINDRUM II \cite{sindrumII} eliminates roughly $35\%$ of the model points for $\Lambda_{IR}\simeq 1.5$\,TeV, implied by EWPM. As for future Muon decay and conversion experiments, it can be seen that a non observation of $\mu\to e$ conversion in DeeMEE \cite{DeeMee} $({\rm Br}(\mu\to e)\lesssim \mathcal{O}(10^{-14})$ already eliminates roughly $93\%$ of the model points. To release this constraint we have to reduce the strength of cross brane interactions in the charged lepton sector, which forces us to re-match the neutrino mixing angles by  gradually pushing the 5D Yukawas in the neutrino sector closer to their perturbativity limit ($|Y|\lesssim 10$). A continued non observation of $\mu\to e$ in MU2e \cite{Mu2e}, COMET \cite{COMET}, PRIME \cite{PRIME} and Project-X \cite{ProX} will require very weak charged lepton ``cross-brane" interactions and thus a too large value for $Y$. Turning off the ``cross-talk" between the UV and IR brane, implies a complete protection from both gauge and fermionic KK mixing effects, as is the case for the earlier brane localized versions of  RS-A$_4$ \cite{Csaki:2008qq,JoseA4}. The latter protection stems from the fact that at LO no RH rotation is necessary $(V_R^\ell=\mathbbm{1})$ and thus one can always find a basis in which the kinetic and mass terms are simultaneously diagonal, which completely forbids tree level FCNC. These conclusions also hold true once we consider the full KK mass matrix since all $3\times3$ blocks mimic the structure of the ZMA mass matrices. In particular, the blocks corresponding to interactions with the custodians $(\tilde{e}_{L,R}^{(1^{+-})},\tilde{\mu}_{L,R}^{(1^{+-})},\tilde{\tau}_{L,R}^{(1^{+-})})$ have the same structure as the Dirac mass matrix (Eq.~(\ref{DiracMassMatrix})), which in turn is also diagonalized by the same rotation $V_{L,R}^{\nu^D}=V_L^\ell=U(\omega)$. Thus, the RS-A$_4$ setup is interesting for cLFV, due to the non-custodial protection mechanism of the $Z$ coupling matrices by virtue of the leading order ``form diagonalizability", which survives to a certain extent when cross-brane interactions are turned on. 

To summarize, we wish to stress that while anarchy is still viable in explaining the neutrino mixing pattern, we have chosen here the approach of looking for an underlying structure that can account for non trivial correlations among observables. It remains to be seen what future experimental data will tell us. Moreover, anarchic models generally imply a larger amount of flavor violation and are thus subject to stronger indirect constraints. Rather inevitably, the most common protection mechanisms developed to relax these constraints, need again to invoke a certain degree of underlying flavor symmetries.

\bibliography{RS-A4_LFV_B}
\bibliographystyle{utphys}

\end{document}